\documentclass[aip, jcp
 sd, noshowpacs,
 amsmath,amssymb,
 reprint,
]{revtex4-1}

\usepackage{graphicx}%
\usepackage{dcolumn}%
\usepackage{bm}%
\begin{document}

\preprint{AIP/123-QED}
\title[]{
Microcanonical rates from ring-polymer molecular dynamics: 
Direct-shooting, stationary-phase, and maximum-entropy approaches 
}

\author{Xuecheng Tao}
\author{Philip Shushkov}
\author{Thomas F. Miller III}
\thanks{Electronic mail: tfm@caltech.edu.}
\affiliation{
Division of Chemistry and Chemical Engineering, California Institute of Technology,\\ Pasadena,
California 91125, USA 
}

\date{\today}

\begin{abstract}
We address the calculation of microcanonical reaction rates for processes involving significant nuclear quantum effects using ring-polymer molecular dynamics (RPMD), both with and without electronically non-adiabatic transitions.
After illustrating the shortcomings of the naive free-particle direct-shooting method, in which the temperature of the internal ring-polymer modes is set to the translational energy scale,
we investigate alternative strategies based on the expression for the  microcanonical rate in terms of the inverse Laplace transform of the thermal reaction rate.
It is  shown that simple application of the stationary-phase approximation (SPA) dramatically improves the performance of the microcanonical rates using RPMD, particularly in the low-energy region where tunneling dominates.   Using the SPA as a Bayesian prior, numerically exact RPMD microcanonical rates are then  obtained using maximum entropy inversion of the thermal reaction rates, for both electronically adiabatic and non-adiabatic model systems.  
Finally, the direct-shooting method is revisited using the SPA-determined temperature for the internal ring-polymer modes, leading to a simple, direct-simulation method with  improved accuracy in the tunneling regime.
\end{abstract}

\pacs{Valid PACS appear here}%
\maketitle

\section{Introduction} 
Ring-polymer molecular dynamics (RPMD) \cite{craig2004quantum, habershon2013ring} has proven to be a useful tool for the calculation of chemical reaction rates, \cite{craig2005chemical, craig2005refined} spectra, \cite{habershon2008comparison, rossi2014remove} and transport coefficients. \cite{miller2005quantum, miller2005quantumwater} The method has been widely applied for the study electronically adiabatic processes for which nuclear quantum effects play an important role, \cite{jiang2019imaging, wilkins2017nuclear, rossi2016nuclear, ceriotti2013nuclear, marsalek2016ab, menzeleev2011direct, menzeleev2010ring, kretchmer2013direct, kretchmer2015tipping, suleimanov2016chemical, collepardo2009bimolecular, perez2012chemical, li2012ring, li2013rate,  suleimanov2012surface, boekelheide2011dynamics, miller2008isomorphic, kreis2017classical} and extensions of the method for systems involving electronically non-adiabatic processes are increasingly common. \cite{hele2005an, duke2017mean, menzeleev2014kinetically, kretchmer2017kinetically, kretchmer2018fluctuating, ananth2013mapping, pierre2017mapping, richardson2013communication, chowdhury2017coherent, tao2018path} 
However, despite the utility of RPMD for calculating quantities in terms of thermal transport coefficients, less work has focused on the extension of the method to non-thermal initial distributions \cite{welsch2016non} 
 or for the calculation of properties associated with non-thermal ensembles, such as microcanonical reaction rates, which would be of use for both benchmarking and practical applications.

Application of RPMD beyond the canonical ensemble immediately encounters the question of how to treat the temperature associated with the intra-bead ring-polymer potential. This temperature is well-defined in thermal applications
for which RPMD was initially developed,\cite{craig2004quantum,habershon2013ring} and it has been justified for RPMD with particular non-equilibrium initial conditions. \cite{welsch2016non}  
In the context of microcanonical reaction rates,  a  direct-shooting method based on the free-particle temperature has been proposed,\cite{duke2015simulating}
in which the 
internal ring-polymer temperature is fixed based on the microcanonical energy, i.e. $ T = E / k_{\rm B}$. 
This protocol has been employed in several model calculations, \cite{duke2015simulating, shakib2017ring} although its reliability has not been systematically examined.

The current work addresses the challenge of microcanonical rate calculations using RPMD. 
In addition to analyzing the previously proposed free-particle direct-shooting protocol, we introduce alternative stationary-phase and maximum-entropy inversion methods to extract microcanonical rates from thermal reaction rates, the calculation of which is well established using RPMD. 
Finally, we return to the direct-shooting method for microcanonical RPMD rates, replacing the free-particle temperature with the optimal temperature from the stationary-phase inversion, which is shown to yield greatly improved microcanonical RPMD rates in the low-energy regime.
Numerical examples of these microcanonical RPMD methods are presented for both electronically adiabatic and non-adiabatic systems.

\section{Methods}

\subsection{Thermal reaction rates from RPMD}
We begin by briefly reviewing RPMD and its use for the calculation of thermal reaction rates.  The theory is presented for a one-dimensional system, and extension to multiple dimensions is straightforward. 
Consider an electronically adiabatic system with the Hamiltonian 
\begin{align} \label{h1l}
\hat{H}=\frac{\hat{p}^2}{2m}+ V (\hat{q}),
\end{align}
where $V (\hat{q})$ is the potential energy function.  
Expressing the  quantum canonical partition function, $Q$, in the path-integral representation yields\cite{feynman1965quantum,chandler1981exploiting,parrinello1984study} 
\begin{align}
Q  &= {\rm tr} \left[ e^{-\beta \hat{H}} \right] \nonumber \\
&= \lim_{n\rightarrow\infty}
\left(\frac{n}{2\pi \hbar}\right)^n \hspace{-3pt} \int \hspace{-1pt}  d \textbf{p} \hspace{1pt} d\textbf{q} \hspace{2pt} e^{-\beta H_n^{\rm iso} \left(\textbf{p}, \textbf{q} \right)},\label{Qpi}
\end{align}
where $\beta$ and $n$ are the reciprocal temperature and the number of imaginary time discretization steps, respectively;
$\textbf{q}=\{q_1, q_2, \ldots, q_n\}$ denotes the positions of the ring-polymer beads, and  $\textbf{p}$ denotes the corresponding momenta. 
Eq.~\ref{Qpi} introduces the classical isomorphic ring-polymer Hamiltonian,
\begin{align}  \label{rpmdH}
H_n^{\rm iso}(\textbf{p}, \textbf{q}) = 
&\sum_{\alpha=1}^n \frac{p_{\alpha}^2}{2m_n}  
+ U_{\rm spr} (\textbf{q}) 
+  \frac1n \sum_{\alpha=1}^n V(q_{\alpha}),
\end{align} 
with $\beta_n =\beta/n$, $m_n = m/n$, and neighboring ring-polymer beads are connected via harmonic springs
\begin{align}
U_{\rm spr} (\textbf{q}) =
\frac12 \hspace{2pt} \frac{m_n}{\beta_n^2} \, \sum_{\alpha=1}^n   \left( q_{\alpha} - q_{\alpha+1} \right)^2.
\end{align} 
Classical sampling of the ring-polymer Hamiltonian  faithfully 
preserves quantum Boltzmann statistics.  The classical equations of motion associated with the ring polymer Hamiltonian are given by 
\begin{align}  \label{rpmd_eom}
 \ddot{q}_\alpha = \frac{1}{\beta_n^2} (q_{\alpha+1}+q_{\alpha-1} - 2 q_\alpha)  - \frac{1}{m} \frac{\partial V(q_\alpha)}{\partial q_\alpha}.
\end{align}

The calculation of thermal rates from RPMD then simply follows from the application of  classical rate theory to the dynamics associated with the ring-polymer Hamiltonian.\cite{craig2005chemical, craig2005refined,habershon2013ring}  Specifically, calculation of the thermal RPMD rate in the flux-side formulation yields
\begin{align} \label{rpmdrate1l}
k {Q_r} 
= &  \lim_{n \to \infty } \lim_{t \to ``\infty" }  \left(\frac{n}{2\pi \hbar}\right)^n 
\int \hspace{-2pt} d\textbf{p}_0 d\textbf{q}_0 
\hspace{2pt} e^{-\beta H_n^{\rm iso}(\textbf{p}_0, \textbf{q}_0)}   \nonumber \\
& \times  \delta({\bar{q}_0- q^\ddagger}) \hspace{2pt}
 \bar{v}_0 \hspace{2pt} h({\bar{q}_t- q^\ddagger}),
\end{align}
which correlates the positions and velocities of the ring-polymer beads at time $t$ following evolution according to the ring-polymer equations of motion (Eq.~\ref{rpmd_eom}) from an initial distribution in which the ring-polymer centroid is positioned at the dividing surface for the reaction.  Here, $\bar{q}_0$ and $\bar{q}_t$ indicate the ring-polymer centroid position at time zero at times zero and $t$, respectively, and $\bar{v}_0$ indicates the centroid velocity at time zero.  $Q_r$ denotes the reactant partition function, $q^\ddagger$ indicates the position of the dividing surface that separates the reactant and product, $h$ is the Heaviside function, and $\delta$ is the Dirac delta function.

\subsection{Microcanonical reaction rates from RPMD} \label{sec2b}

In the following, we describe three alternative strategies for calculating microcanonical reaction rates from RPMD.  The first involves an inverse Laplace transform of the thermal RPMD reaction rates and introduces no approximations beyond that of the thermal RPMD rate theory, although it is numerically the most demanding. 
The subsequent two methods introduce additional approximations (i.e., the stationary phase approximation and the direct shooting approximation) with the benefit of reduced numerical complexity.

\subsubsection{Maximum entropy inversion}
Reaction rates in the microcanonical and canonical ensembles are connected via the Laplace transform\cite{miller1975path}
\begin{align} \label{mtc} 
k(\beta) {Q_r}(\beta) 
= \frac1{2\pi \hbar} \int_{-\infty}^{+\infty} \hspace{-1pt} dE \hspace{2pt}  e^{-\beta E}  N(E),
\end{align}
which can be inverted to yield 
\begin{align} \label{ctm}
N(E) 
&= (2\pi \hbar) \hspace{4pt} \frac1{2\pi i} \int_{\gamma-i\infty}^{\gamma+i\infty} 
d\beta  \hspace{2pt}  e^{\Phi (\beta)}
\end{align}
with
\begin{align} \label{phidef}
\Phi (\beta) = 
{\beta E} + \log \left( k {Q_r}  \right).
\end{align}
The line integral in  Eq.~\ref{ctm} is performed along  
${\rm Re}[\beta] \hspace{-2pt}  = \hspace{-2pt}  \gamma$, 
where $\gamma$ is greater than the real part of all points for which $\Phi$ is singular. 
However, numerical implementation of this Laplace transform 
is typically ill-conditioned and  sensitive to statistical noise, \cite{press2007numerical} which is unavoidable in simulation-based  thermal rate calculations.  

To ameliorate this problem, we  employ 
the maximum entropy (MaxEnt) method, \cite{bryan1990maximum, jarrell1996bayesian}  
which utilizes statistic inference and a Bayesian prior to regularize the numerical  inversion.\cite{jarrell1996bayesian, habershon2007quantum, miller2008isomorphic, rabani2000quantum, golosov2003analytic}   
MaxEnt is implemented by rewriting the integral in Eq.~\ref{mtc} in  matrix form,
\begin{align} \label{maxenteq}
\boldsymbol{\kappa} = \mathbf{B} \hspace{2pt} \boldsymbol{\nu},
\end{align} 
where $\boldsymbol{\kappa}$ is the vector of  thermal rate input data at discrete temperature points $\{\beta_i\}$, and $\boldsymbol{\nu}$ is the vector of microcanonical rate outputs at discrete energy values $\{E_j\}$. 
Specifically, the elements of $\boldsymbol{\kappa}$ and $\boldsymbol{\nu}$ are
$\kappa_i = 2\pi \hbar \hspace{2pt} k(\beta_i) Q_r(\beta_i) $ and $\nu_j  = N(E_j) $. 
The matrix $\mathbf{B}$ is comprised of the Boltzmann kernel $B_{ij}=e^{-\beta_i E_j} \Delta E_j$, where $\Delta E_j = E_{j+1} - E_j$ is the integration stepsize. 
MaxEnt yields the microcanonical rate by maximizing the objective function
\begin{align} \label{maxent_objfun}
Q(\boldsymbol{\nu}; \alpha) = \alpha S(\boldsymbol{\nu}) - \chi^2(\boldsymbol{\nu})/2
+ V_{\rm reg}(\boldsymbol{\nu}),
\end{align}
where the information entropy $S$ describes the degree to which solution is faithful to a prior model $\boldsymbol{\lambda} (\{ E_j \})$,
\begin{align}
S(\boldsymbol{\nu})= \sum_j \left(\nu_j -\lambda_j - \nu_j \log \frac{\nu_j}{\lambda_j} \right),
\end{align}
and the likelihood function $\chi^2$ describes the accuracy with which the reference thermal rate data is fit,
\begin{align} \label{chidef}
\chi^2 (\boldsymbol{\nu}) &= (\boldsymbol{\kappa} - \mathbf{B} \boldsymbol{\nu})^\text{T}
\hspace{2pt} \mathbf{C}^{-1}  \hspace{1pt}  (\boldsymbol{\kappa} - \mathbf{B} \boldsymbol{\nu}).
\end{align} 
Here, $\mathbf{C}$ is the covariance matrix for the thermal rate data with  elements
\begin{align}
C_{ii'} = & \delta_{ii'} \sigma_i^2,
\end{align}
where $\delta_{ii'}$ is the Kronecker delta function, and $\sigma_i$ is the standard deviation for the $i$-th thermal rate datapoint. 
The parameter
$\alpha$ balances between  accurately fitting the reference data while preserving the prior model.

Finally,  $V_{\rm reg}(\boldsymbol{\nu})$ penalizes those trial solutions that violate the physical constraints of the microcanonical rate constant, namely that it satisfy 
$N(E) \hspace{-2pt} \in \hspace{-2pt} [0, 1]$.
The lower bound is enforced by conducting a solution search only in the positive subspace
while the upper bound is enforced via the functional form
\begin{align}
V_{\rm reg}(\boldsymbol{\nu}) = - \sum_j \frac12 \hspace{2pt} \zeta \hspace{4pt} I^2( \nu_j - 1 ), 
\label{regpotential}
\end{align}
where 
\begin{align}
I(\nu_j-1)= &
\begin{cases}
\nu_j-1, & \nu_j \ge 1, \\
0, & \nu_j< 1,
\end{cases}
\end{align}
and $\zeta$ is chosen according to a tolerance criterion. 

\subsubsection{Stationary phase approximation} \label{secIIB3}

As an alternative to numerically exact inversion, we apply the stationary phase approximation (SPA)\cite{miller2006applied, miller2003quantum} to Eq.~\ref{ctm}. 
Implementation of the SPA involves finding the stationary point of the phase function $\Phi$ in Eq.~\ref{phidef} and then approximating the integrand as a Gaussian function along the imaginary axis. Setting the first-order derivative of the phase function to zero yields the energy-temperature correspondence \cite{miller2003quantum}
\begin{align} \label{statbeta}
 E_{\rm st} = -  \left. \frac{\partial \log \left[  \hspace{2pt} k(\beta) Q_r (\beta)  \hspace{2pt} \right] }{\partial \beta} \right|_{\beta_{\rm st}}
  , \qquad \beta_{\rm st} \in \mathbb{R}
\end{align}
where $\beta_{\rm st}$ is the stationary temperature that is assumed to dominate the integrand. 
The resulting SPA microcanonical rate prediction is given by
\begin{align} \label{statrate}
N_{\rm SPA} (E_{\rm st}) =  
&\frac{ 2\pi \hbar}{\sqrt{2\pi}} \hspace{2pt}  \left( \left.\frac{\partial^2 
\log \left[  \hspace{2pt} k(\beta) Q_r (\beta)   \hspace{2pt} \right] }{\partial \beta^2 } \right|_{\beta_{\rm st}} \right)^{-1/2} \nonumber \\ 
&\times e^{\beta_{\rm st} E_{\rm st}} \hspace{2pt}  k(\beta_{\rm st}) Q_r (\beta_{\rm st}) .
\end{align}

A well-known shortcoming of the SPA is that the calculated microcanonical rate violates the constraint $N(E) \hspace{-2pt} \leq \hspace{-2pt} 1$ in the high-energy limit.\cite{miller2003quantum}
In this regime, 
the barrier-crossing dynamics reduces to free particle motion, and the thermal rate becomes
\begin{align} \label{fp}
{\left[ k Q_r \right] ^{\rm FP}}= 1/\left( 2 \pi \beta \hbar \right).
\end{align}
Substituting Eq.~\ref{fp} into Eqs.~\ref{statbeta} and  \ref{statrate} yields the  energy-temperature correspondence relation
 \begin{align}
E_{\rm st}^{\rm FP} = 1 / \beta_{\rm st}^{\rm FP},
\end{align}
and the corresponding microcanonical rate in the high-energy limit is 
\begin{align}
\label{wronglimit}
N_{\rm SPA}^{\rm FP}(E) = e / \sqrt{2 \pi} \simeq 1.084,
\end{align}
in excess of the correct upper limit. 

\subsubsection{Direct shooting approximation}
By analogy with classical rate theory, 
a physically intuitive  strategy for approximating microcanonical rates from RPMD is to simply 
(\emph{i}) initialize trajectories from the reactant side with specified translational energy, 
(\emph{ii}) propagate those trajectories using the microcanonical equations of motion in Eq.~\ref{rpmd_eom}, and (\emph{iii}) count the proportion of trajectories that reach the product region, 
such that
\begin{align}  \label{direct}
N_{\rm direct}&(E,  \beta_{\rm int}) \nonumber \\ 
 =&  \lim_{n \to \infty }  \lim_{t \to ``\infty" }  
\hspace{1pt} \frac{n^n}{\left(2\pi \hbar\right)^{n-1}} 
\hspace{-3pt} \int \hspace{-2pt} d\textbf{p}_0 d\textbf{q}_0 
 \frac{ e^{-\beta_{\rm int} H_n^{\rm iso}(\textbf{p}_0, \textbf{q}_0)}}{ e^{-\beta_{\rm int} E}}  \nonumber \\
&\times {\delta \left[ {\bar{p}_0 \hspace{-1pt} - \hspace{-1pt} \sqrt{2m(E-V_a)}} \right] }
 \hspace{2pt} 
 \delta({\bar{q}_0 \hspace{-1pt} - \hspace{-1pt} q^\ddagger}) \hspace{2pt}  h({\bar{q}_t \hspace{-1pt} - \hspace{-1pt} q^\ddagger}),
\end{align}
where $\bar{p}=\sum_\alpha p_\alpha$ is the centroid momentum, and $V_a$ is the potential energy in the reactant asymptotic region. 
The centroid kinetic energy for the RPMD trajectories are initialized to match the physical incident energy (as indicated by the $\delta$ function), while the internal modes are thermally sampled from an internal temperature $\beta_{\rm int}$, the appropriate value of which is not obvious.
Previously,\cite{duke2015simulating} the direct shooting method for calculating microcanonical rates has been applied with the internal temperature set in correspondence to the physical incident energy $\beta_{\rm int} = 1/ E$, which we call the free-particle protocol; in the current work, we shall also consider a prescription for the internal temperature that is derived from the SPA. We note that  direct shooting  is similar in practical implementation to the calculation of non-equilibrium time-correlation functions using RPMD with momentum-kick initial conditions,\cite{welsch2016non, jiang2019imaging} although the theoretical justification is more clearly established for the case of non-equilibrium time-correlation functions than  for the calculation of microcanonical rates as considered here.

\section{Computational Details}

Unless specified, all results are reported in atomic units. 

Implementation of the direct-shooting approach for microcanonical rates  employs Eq.~\ref{direct}. Initial configurations for the ring polymer are sampled from the thermal distribution associated with the  specified internal temperature ($\beta_{\rm int}$), while dynamical evolution is performed using the standard RPMD integration scheme with a timestep of $0.3$. This choice of  timestep is confirmed to avoid resonance instabilities, although we note that the best practice for future applications is to employ the Cayley-modification to the RPMD integration.\cite{korol2019cayley,korol2019cayley2} Calculations with up to $144$ ring-polymer beads are performed to ensure the convergence of the path-integral discretization.

Implementation of the SPA utilizes Eqs.~\ref{statbeta} and \ref{statrate}.
Eq.~\ref{statbeta} is first solved to obtain the stationary temperature from the thermal rate data. 
Then, the SPA microcanonical rate is  obtained using Eq.~\ref{statrate}.  
First- and second-order derivatives of $k Q_r$ are obtained from a standard basis-spline interpolation procedure. \cite{dierckx1995curve, scipy} 
Validation of the numerical procedure is performed by comparison with independent SPA results obtained from path-integral Monte Carlo sampling methods (Appendix A).

Implementation of the MaxEnt approach closely follows the Bryan algorithm. \cite{bryan1990maximum, jarrell1996bayesian} Calculations are performed with a modified version of an open-source code.\cite{code} 
Quantum mechanical and RPMD thermal rates are the fitting targets in these calculations,
while SPA microcanonical rates are employed as the Bayesian prior.
While not included here, results were also obtained using the flat Bayesian prior,  $\boldsymbol{\lambda}(E)=1$; however, the numerics of these calculations were found to be less stable than those based on the SPA, which requires no additional information beyond the thermal rates that are also used for the fitting target.
To ensure that the SPA priors are nonzero and sufficiently smooth, they 
are filtered with a low threshold value of $10^{-3}$, followed by a simple moving average procedure to suppress the statistical fluctuations. 
A regulation potential (Eq.~\ref{regpotential}) with $\zeta=10^7$ is sufficient to enforce the upper bound of $N(E)$
to a tolerance of $10^{-5}$
in all reported calculations. 

\begin{figure}[t]
\centering
\includegraphics[width=0.95\columnwidth]{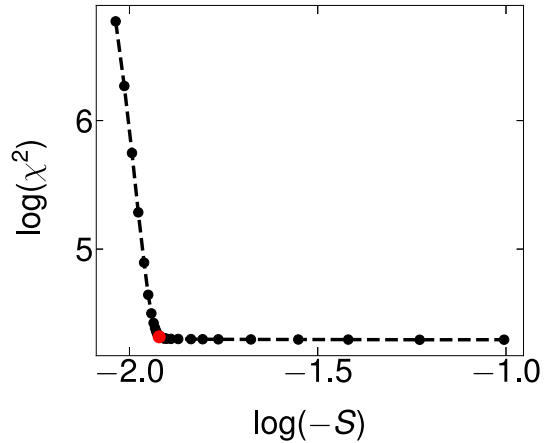}
\caption{\label{fig4a} 
An illustrative example of the `L-curve' that is used to determine the parameter $\alpha$ in each MaxEnt calculation. The optimal value of $\alpha$ coincides with the kink in the curve (indicated by a red point).  This example  corresponds to the Eckart barrier, using RPMD thermal rate data for the fitting target and the SPA-RPMD microcanonical rates for the Bayesian prior. 
}
\end{figure}

To specify the parameter $\alpha$ in the objective function of the MaxEnt calculations (Eq.~\ref{maxent_objfun}), the `L-curve' rule was employed as in many previous studies. \cite{rabani2000quantum, habershon2007quantum, miller2008isomorphic}  
As illustrated with a representative example in Fig.~\ref{fig4a}, 
the balance between fitting accuracy and solution likelihood when plotted as a parametric function of $\alpha$ yields a hockeystick-shaped curve.  We take the kink of the curve (red point) to correspond to the optimal balance between these attributes.

\section{Results}
\subsection{Microcanonical RPMD rates for electronically adiabatic reactions}

We begin by analyzing the effectiveness of the direct shooting approach with different choices of ring-polymer internal temperatures, $\beta_{\rm int}$. The symmetric Eckart barrier model for H+H$_2$ reactive scattering \cite{craig2005refined} is chosen as the test example, with potential energy function
\begin{align}
V(q) = {V_0} \left/ {\cosh^2 (q/q_0)} \right.
\end{align}
using parameters $m=1061$, 
$V_0=0.425$ eV, and $q_0=0.734$. 
Analytical solution of the microcanonical rate for this model yields 
\begin{align} \label{eckqm}
N(E) = &  f / \left( f+g \right),\ \textrm{where} \nonumber \\
f = &\sinh^2 \left( \pi q_0 \sqrt{2mE}/ \hbar \right),\ \textrm{and} \nonumber \\
g = &\cosh^2 \left( \pi \sqrt{ \left| 2mV_0q_0^2 / \hbar^2 -1 / 4 \right|} \right),
\end{align}
and the exact thermal rate is obtained by integrating $N(E)$ over Boltzmann kernel, following Eq.~\ref{mtc}. 

\subsubsection{Free-particle direct shooting}

\begin{figure}[tb]
\centering
\includegraphics[width=\columnwidth]{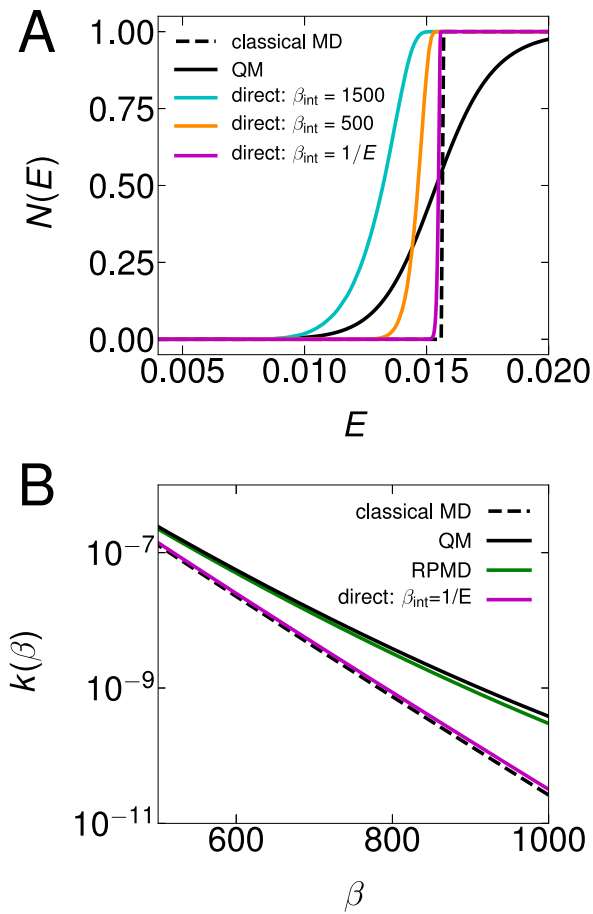}
\caption{\label{fig1} 
{\bf{(A)}} Microcanonical rate predictions for the 
Eckart barrier. Results are calculated with classical mechanics (classical MD, dashed black), analytical quantum mechanics (QM, solid black) and direct shooting approach with different internal ring-polymer temperatures (direct, solid cyan, orange and magenta). The free-particle direct shooting protocol is labeled $\beta_{\textrm{int}}=1/E$. {\bf{(B)}} Thermal rates obtained by substituting  microcanonical rates from 
various levels of theory (classical MD, QM, and free-particle direct shooting) into Eq. \ref{mtc}. For comparison, the standard RPMD thermal rates (green) are also included. 
}
\end{figure}

Fig.~\ref{fig1}A plots the microcanonical rate prediction from classical mechanics, exact quantum mechanics, and the direct shooting scheme (Eq.~\ref{direct}), as a function of energy.
As expected, the step-function shape of the classical result is smoothed 
due to nuclear quantum effects. It is clear from the figure that the 
direct-shooting scheme is strongly sensitive to the choice of internal ring-polymer temperature, particularly at low temperatures for which tunneling plays an important role; irrespective of the employed value of $\beta_{\rm int}$, the direct-shooting scheme reverts to classical behavior in the high-energy regime.
Strikingly, almost all nuclear quantum effects are absent using the free-particle protocol ($\beta_{\rm int} = 1/ E$) for the internal ring-polymer temperature.

Fig.~\ref{fig1}B presents the canonical reaction rates for the Eckart barrier as a function of temperature, including exact quantum and classical results, as well as the standard RPMD calculation of thermal reaction rate (green).\cite{craig2005chemical, craig2005refined}
As is well known for such problems, RPMD allows for the direct calculation of thermal reaction rates with good accuracy.
However, the figure also shows the results of the RPMD thermal rate prediction obtained by transforming (via Eq.~\ref{mtc}) the microcanonical RPMD rates from the free-particle direct-shooting protocol (magenta).  Consistent with the results of Fig.~\ref{fig1}A, this direct-shooting protocol provides an essentially classical description of the thermal reaction rate across the entire range of temperatures.  Fig.~\ref{fig1}B clearly demonstrates that approximation of microcanonical rates via the free-particle direct-shooting method (magenta vs. solid black) is a far greater source of error than the intrinsic approximation of RPMD for calculating thermal rates (green vs. solid black).  This figure illustrates the hazards of using direct shooting for RPMD microcanonical rates, and it suggests that better results for microcanonical rate should be achievable on the basis of RPMD dynamics. 

\subsubsection{Stationary phase approximation}

\begin{figure}[!htbp]
\centering
\includegraphics[width=0.95\columnwidth]{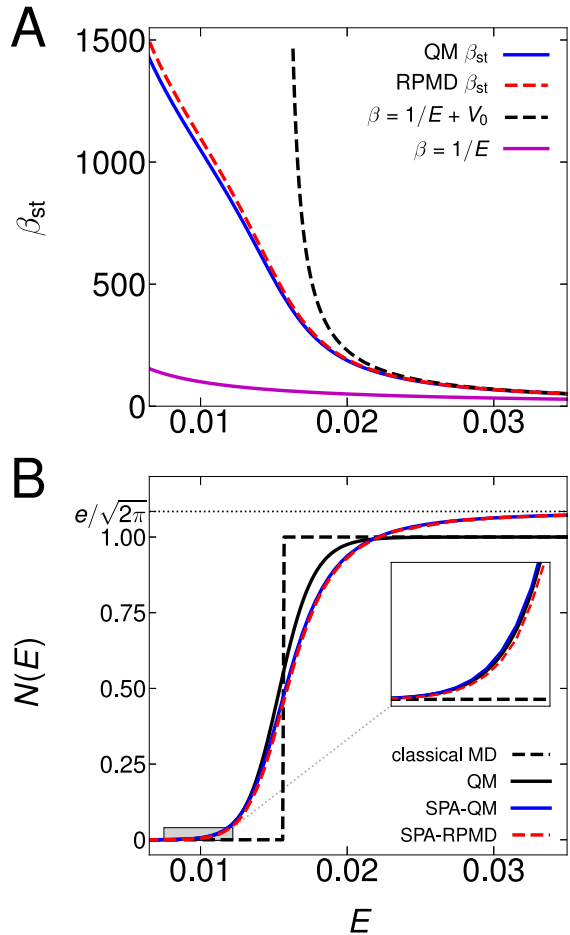}
\caption{\label{fig2} 
Stationary phase approximation (SPA) results for the  Eckart barrier.
{\bf{(A)}} The stationary temperature calculated with QM (blue) and RPMD (dashed red) thermal rates, respectively, as a function of incident energy. For comparision, the classical asymptote of the stationary temperature (dashed black), and the free-particle temperature (magenta) are also included. {\bf{(B)}} Microcanonical rates obtained using the SPA with input from QM (blue) and RPMD (dashed red) thermal rates, respectively.  For comparison, the  microcanonical rates from exact QM and classical MD are also included.  The inset expands the low-energy regime.
}
\end{figure}

We now turn our attention to the use of the SPA to calculate RPMD microcanonical reaction rates for the example of the Eckart barrier.
Fig.~\ref{fig2}A presents the calculated stationary temperature $\beta_{\textrm{st}}$ as a function of  energy, obtained via   Eq.~\ref{statbeta} with input from either exact quantum thermal rates (blue) or standard RPMD thermal rate calculations (red, dashed).  For comparison, we also show the temperature associated with the  free-particle protocol ($\beta=1/E$), which differs significantly from the stationary temperature at each energy, 
as well as an analytical expression for the high-energy stationary temperature ($\beta=1/E + V_0$) that is derived in Appendix A. 
In the low-energy regime, only the RPMD thermal rate data provides a satisfactory description of the stationary temperature obtained from the exact quantum results.

Fig.~\ref{fig2}B presents  microcanonical rates from the SPA (Eq. \ref{statrate}) using input from RPMD thermal rates. 
To provide a baseline of accuracy associated with the SPA, we first compare the microcanonical rate from exact quantum mechanics (black, solid) with that obtained via SPA applied to exact quantum thermal rates (blue).  The difference in these curves presents a best-case scenario for the accuracy of a method that approximates microcanonical rates via SPA; and it is seen that while the agreement at low temperature is excellent, there is substantial deviation associated with energies in the high-energy regime. Encouragingly, essentially identical performance is seen when the microcanonical rates are obtained via application of the SPA to  RPMD thermal rates (red, dashed). This indicates that the RPMD thermal rates are a  smaller source of error than the SPA, particularly in the high-energy regime.  Finally, comparison of the results in Fig.~\ref{fig2}B with Fig.~\ref{fig1}A makes clear that RPMD offers a much more accurate avenue to the calculation of microcanonical rates than might be concluded from simulations that employ direct shooting.

We again note that the SPA errors at high energy in Fig.~\ref{fig2}B that are well known and due to the neglect of higher-order terms in the phase function.\cite{miller2003quantum}
As anticipated form Eq.~\ref{wronglimit}, both sets of SPA results in the figure converge to the  erroneous high-energy asymptote of $1.084$.

\subsubsection{Maximum entropy inversion}

\begin{figure}[!htbp]
\centering
\includegraphics[width=0.95\columnwidth]{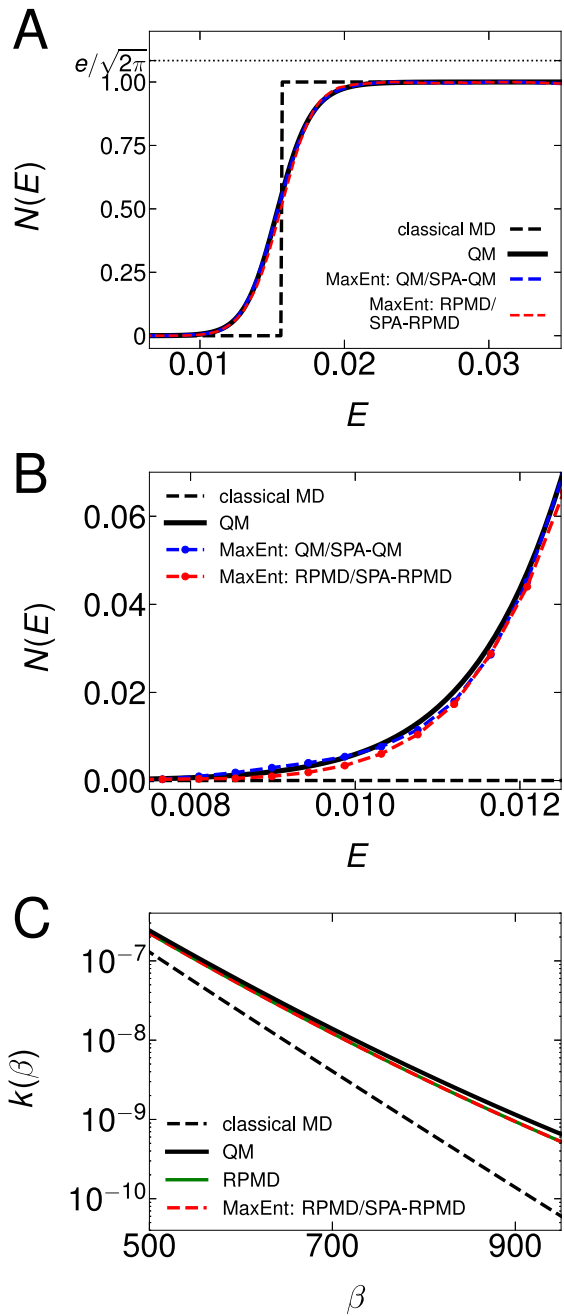}
\caption{\label{fig3} 
Maximum entropy (MaxEnt) inversion results for the  Eckart barrier. 
Methods are labeled with the format `MaxEnt-[thermal rate input type]/[prior type]'.
{\bf{(A)}} 
MaxEnt solutions for the microcanonical reaction rate as a function of incident energy.  
Microcanonical rates from classical MD and exact QM are also presented for reference.
{\bf{(B)}}  An expanded view of  panel A in the low-energy regime. 
{\bf{(C)}} Thermal rates obtained by integrating $N(E)$ in panel A over the Boltzmann kernel. For comparison, exact QM, classical MD, and standard RPMD thermal rates are also included. 
}
\end{figure}

As the third alternative  for obtaining microcanonical rates from RPMD, Fig.~\ref{fig3}A presents results for the Eckart barrier obtained using MaxEnt inversion.
To establish the baseline error for the MaxEnt procedure, the dashed blue curve presents the results obtained via inversion of the exact QM thermal rates  using the SPA-QM microcanonical rates (Fig.~\ref{fig2}B) as the Bayesian prior.
Finally, the dashed red curve presents the MaxEnt results obtained via inversion of the RPMD thermal rates  using the SPA-RPMD microcanonical rates as the Bayesian prior.  This last result utilizes input from RPMD thermal rates alone.

It is clear from Fig.~\ref{fig3}A that the MaxEnt procedure provides excellent accuracy across the entire range of energies, avoiding the incorrect high-energy asymptote of the SPA results.    Closer examination of the low-energy regime in Fig.~\ref{fig3}B reveal that the  agreement persists even in the regime of strong tunneling.  Comparison of the red curve in Fig.~\ref{fig3}B (MaxEnt:RPMD/SPA-RPMD) with the blue curve in the inset of Fig.~\ref{fig2}B (SPA-QM) suggests that the RPMD thermal rate data is slightly greater source of error than the SPA in the low-energy regime for the Eckart barrier, although all of the differences are small. 
Taken together, the results in Fig.~\ref{fig3}A and B indicate that for this example, the use of MaxEnt inversion helps to improve the quality of the SPA at intermediate and higher energies, but it does little to improve the quality of the SPA in the low-energy regime.

Finally, as a self-consistency check, Fig.~\ref{fig3}C presents the thermal rates obtained by transforming (via Eq.~\ref{mtc}) the microcanonical  rates obtained from the MaxEnt inversion of the RPMD thermal rates (red, dashed).
For comparison, the exact quantum, classical, and standard RPMD thermal rates are also included.  As expected, the MaxEnt RPMD rates are fully consistent with the standard RPMD thermal rates, and both are in good agreement with the exact QM results.  

\subsubsection{Stationary-temperature direct shooting} \label{sec3e}

\begin{figure}[!b]
\centering
\includegraphics[width=0.95\columnwidth]{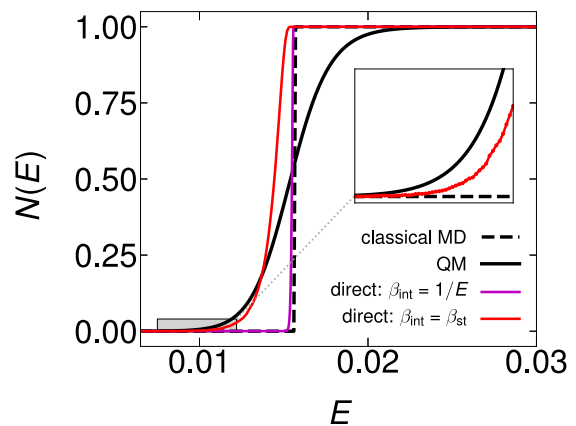}
\caption{\label{fig5}
Microcanonical rate predictions for the  Eckart barrier, comparing the direct-shooting method with the ring-polymer internal temperature set to either the stationary temperature (red) or the free-particle temperature (magenta).  Also included are the exact QM and classical MD results.
}
\end{figure}

Given the success of the SPA for extracting microcanonical rates from standard RPMD thermal rates, it is tempting to see whether data obtained from the SPA can be used to improve the direct shooting method. Specifically, we explore the use of the stationary temperature as the 
ring-polymer internal temperature for initializing and propagating the direct-shooting trajectories, i.e., seting $\beta_{\rm int} = \beta_{\rm st} $ in Eq.~\ref{direct}. 
This strategy is physically appealing, since the
stationary temperature (which is a function of incident energy, see Fig.~\ref{fig2}A) dictates the delocalizaton of the ring-polymer  in its barrier-crossing configuration;\cite{miller2003quantum,richardson2009ring,richardson2017microcanonical} 
also note that the stationary temperature approaches the  free-particle temperature at high incident energies. 

Fig.~\ref{fig5}
plots the microcanonical rate for the Eckart barrier, obtained using  the stationary-temperature direct-shooting method (red).  For comparison, the free-particle direct-shooting (magenta), classical MD (black, dashed), and exact quantum (black, solid) results are all reproduced from Fig.~\ref{fig1}A.
While  stationary-temperature direct shooting  remains qualitatively less accurate than the SPA and MaxEnt inversion methods, it nonetheless substantially improves the results of the free-particle direct-shooting approach in the low-energy region where tunneling is important.  
These results indicate that stationary-temperature direct shooting is a less quantitative tool than SPA or MaxEnt for the calculation of microcanonical rates from RPMD trajectories, but it may nonetheless prove useful in exploratory studies for which a direct trajectory-based simulation approach is needed, or in applications to high-dimensional systems for which obtaining precise thermal reaction rate in the whole temperature region is computationally expensive.

\subsection{Microcanonical RPMD rates for non-adiabatic systems}

Although we have thus far only discussed the SPA and MaxEnt inversion methods in the context of single-level (i.e., electronically adiabatic) processes, both methods can be naturally extended to multi-level systems. 
Specifically, given state-resolved thermal reaction rates for a non-adiabatic process, both the SPA and MaxEnt methods can be used to compute state-resolved microcanonical rates for different reaction channels. 

For the SPA method, state-resolved thermal reaction rates are substituted into Eqs. \ref{statbeta} - \ref{statrate}, respectively, yielding a single stationary temperature and a single state-resolved  microcanonical rate for each reaction channel.
For the MaxEnt method, we solve the 
coupled integral equation
\begin{align}
\left(
\begin{array}{c}
\boldsymbol{\kappa}_{1\to2}  \\ 
\boldsymbol{\kappa}_{2\to2} \\  
\end{array}
\right) 
=
\left(
\begin{array}{cc}
\mathbf{B} &  \\ 
  & \mathbf{B} \\  
\end{array}
\right) \cdot
\left(
\begin{array}{c}
\boldsymbol{\nu}_{1\to2}  \\ 
\boldsymbol{\nu}_{2\to2} \\  
\end{array}
\right) 
\end{align}
for a system with two reaction channels (e.g., diabat 1 to 2, and diabat 2 to 2), with $\boldsymbol{\kappa}$, $\mathbf{B}$ and $\boldsymbol{\nu}$ defined  in Eq.~\ref{maxenteq}. 
The MaxEnt objective function for the two-level system is
\begin{align}  \label{maxentnon}
Q(\boldsymbol{\nu}_{1\to2}, \boldsymbol{\nu}_{2\to2}; \alpha) = &
\hspace{12pt} 
\alpha S(\boldsymbol{\nu}_{1\to2}) - \chi^2(\boldsymbol{\nu}_{1\to2})/2 \nonumber \\
& + \alpha S(\boldsymbol{\nu}_{2\to2}) - \chi^2(\boldsymbol{\nu}_{2\to2}) /2   \nonumber \\
& + V_{\rm reg}(\boldsymbol{\nu}_{1\to2}, \boldsymbol{\nu}_{2\to2})
\end{align}
which sums the information entropy and likelihood function contributions for the state-resolved rates.
The regularization potential is likewise generalized, 
\begin{align}
V_{\rm reg}(\boldsymbol{\nu}) = - \sum_j \frac12 \hspace{2pt} \zeta \hspace{4pt} I^2( \nu_{1\to2, j} + \nu_{2\to2, j} - 1 ),
\end{align}
to enforce unitarity 
\begin{align}
N_{1\to2}(E_j) + N_{2\to2}(E_j) \le 1  \hspace{10pt} \forall j
\end{align}
The constraint of non-negativity
\begin{align}
N_{1\to2}(E_j) \ge 0, N_{2\to2}(E_j) \ge 0, \hspace{10pt} \forall j,
\end{align}
is enforced as before by confining the solution search to the positive subspace.

\begin{figure}[!tb]
\centering
\includegraphics[width=0.97\columnwidth]{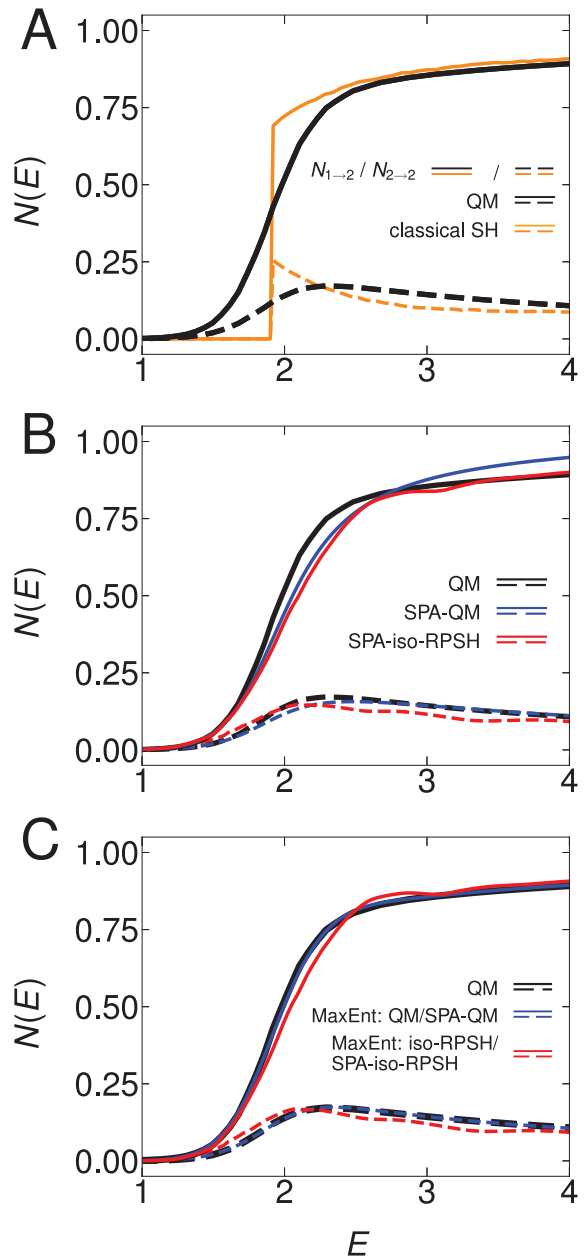}
\caption{\label{fig4} 
State-resolved microcanonical rates for the two-level model in Eq. \ref{scatpot}. 
Dashed lines indicate the 1$\to 2$ diabatic reaction channel and solid lines indicate the 2$\to 2$ diabatic reaction channel. 
{\bf{(A)}} Microcanonical rates from classical surface hopping (orange) and numerically exact QM wavepacket propagation (black). 
{\bf{(B)}} SPA results for the microcanonical rate, with input from exact QM (blue) and iso-RPSH  (red) thermal rates. 
{\bf{(C)}} MaxEnt results for the microcanonical rate, with input from exact QM (blue) and iso-RPSH  (red) thermal rates. 
}
\end{figure}

\begin{table}[tb]
\caption{\label{tab1} Parameters
for the two-level
model in Eq.~\ref{scatpot}.}
\begin{ruledtabular}
\begin{tabular}{cccc}
Parameter&\mbox{Value}&Parameter&\mbox{Value} \\
\hline
$A_1$ & $7$ & $a_1$ & $1$ \\
$A_2$ & $-18/\pi$ & $a_2$ & $\sqrt{3\pi}/4$ \\
$A_3$ & $0.25$ & $a_3$ & $0.25$ \\
$B_1$ & $-0.75$ & $q_1$ & $-1.6$ \\
$B_2$ & $54 / \pi$ & $q_3$ & $-2.625$ \\
\end{tabular}
\end{ruledtabular}
\end{table}

As a numerical demonstration for  non-adiabatic reaction dynamics, 
we use a two-state gas-phase reactive scattering model that has been previously introduced.\cite{tao2018path, tao2019simple}  
In the diabatic representation, the  potential energy functions for this system are
\begin{align}
\label{scatpot}
V_{11}(q)&=\frac{A_1}{1+e^{-a_1\left(q-q_1\right)}}+B_1\nonumber\\
V_{22}(q)&=\frac{A_2}{1+e^{-a_2q}}+\frac{B_2}{4\ {\rm cosh}^2\left({a_2q / 2}\right)}\nonumber\\
V_{12}(q)&=V_{21}(q)=A_3e^{-a_3\left(q-q_3\right)^2}
\end{align}
with parameters specified in Table~\ref{tab1} and with reactant and product regions corresponding to $q \hspace{-2pt}\rightarrow \hspace{-2pt} -\infty $ and  $q \hspace{-2pt} \rightarrow \hspace{-2pt} \infty $, respectively.

We focus on microcanonical rates in the range of incident energies for which the higher-energy state is unavailable as a product channel (i.e., the only two available reactions channels correspond to the $1 \hspace{-2pt}\to \hspace{-2pt} 2$ and $2 \hspace{-2pt}\to \hspace{-2pt} 2$ processes on the diabatic states). 
Calculation of state-resolved thermal rates for the two channels is performed with the flux-side formulation\cite{tao2019simple}  of iso-RPSH,\cite{tao2018path} with both the methodological details and thermal-rate results both reported elsewhere.\cite{tao2019simple}
The only difference in the current work is that 1000-fold more trajectories are performed to suppress statistical error in the thermal rates for the MaxEnt calculations, although the thermal rate results are graphically indistinguishable from those previously published.\cite{tao2019simple}

For comparison, Fig.~\ref{fig4}A presents the state-resolved microcanonical rates obtained from numerically exact quantum mechanics \cite{feit1982solution, neuhauser1991application} and using classical surface hopping\cite{tully1990molecular} as implemented in our previous work.\cite{tao2018path}  
As for the one-level system, the microcanonical rates with classical nuclear dynamics exhibit a sharp increase when the incident energy reaches the barrier height.
Although  classical surface hopping qualitatively includes the effect of the non-adiabatic transition and performs well in the high-energy regime, it fails to capture the significant nuclear quantum effects in this problem.

Fig.~\ref{fig4}B presents the microcanonical rates obtained via application of the SPA to  state-resolved thermal rates from exact QM (blue) and from iso-RPSH (red). Comparison of the QM and SPA-QM results  indicate that the SPA approximation is  a good approximation in this example.  Furthermore, comparison of these curves with the  SPA-iso-RPSH results indicates that the  iso-RPSH thermal rate data is an even smaller source of error than the SPA.
At higher energies, the SPA results exhibit the same pathologies as those discussed in connection with Eq.~\ref{wronglimit}, and the better performance of SPA-iso-RPSH versus SPA-QM in this regime is likely due to fortuitous error cancellation.

Fig.~\ref{fig4}C presents the microcanonical rates obtained via application of the MaxEnt method to the state-resolved thermal rates.  As for the one-level system described in Fig.~\ref{fig3}, the MaxEnt method improves the description for the two-level system at high energies but does little to refine the description of the SPA at lower energies.

Taken together, these results indicate that the iso-RPSH method  can be staightforwardly extended for the accurate calculation of state-resolved microcanonical rates. Moreover, these results show that iso-RPSH provides an accurate description of both the thermal and microcanonical reaction rates of this system in a regime for which both non-adiabatic and nuclear quantum effects play an important role, although the method has been shown to underestimate the asymmetry in the Marcus inverted regime
in the  golden-rule limit of electron transfer.\cite{lawrence2019analysis}

\section{Summary}

Whereas the ring-polymer molecular dynamics (RPMD) thermal rate theory has proven immensely successful in many chemical application domains,  far less attention has been paid to the  problem of calculating microcanonical reaction rates using RPMD, which may be of considerable value in the context of gas-phase and surface-molecule scattering processes.  The current work addresses this shortcoming by exploring a variety of strategies to calculating microcanonical reaction rates with RPMD.
It is found that the ad hoc strategy of direct shooting of RPMD trajectories is strongly sensitive to the internal ring-polymer temperature that is employed; this is somewhat ameliorated in the tunneling regime via the use of an internal temperature based on the stationary-phase approximation (SPA), but the resulting direct-shooting results remain overly classical in the barrier-crossing energy regime.  
Far more accurate microcanonical rates are obtained from 
RPMD thermal rate data via Laplace transform inversion using either the SPA or the numerically exact maximum entropy method.  In general, we find that the SPA applied to RPMD thermal rate data provides the best compromise between good accuracy and numerical feasibility, particularly in the low-energy tunneling regime, although we point out that the alternative direct-shooting and maximum entropy methods described here may also prove useful in particular application cases. 

While the current paper focuses only on the calculation of microcanonical rates from RPMD thermal rate data, we note that
similar strategies can also be applied for the calculation of other time correlation functions, spectra, and transport coefficients in the microcanonical ensemble.

\begin{acknowledgments}
We acknowledge support from the 
Department of Energy under Award No.~DE-FOA-0001912 and the National Science Foundation under  Award
CHE-1611581. 
Additionally, P.S. acknowledges a German Research Foundation (DFG) Postdoctoral Fellowship.
Computational resources were provided by the National Energy Research Scientific Computing Center, which is supported by the Office of Science of the US Department of Energy under Contract  No. DE-AC02-05CH11231.
\end{acknowledgments}

\begin{appendix}

\section{Evaluation of the stationary temperature via path-integral Monte Carlo sampling methods}

To evaluate the stationary temperature $\beta_{\rm st}$ in Eq.~(\ref{statbeta}) via path-integral Monte Carlo sampling methods, we apply the quantum transition state theory (QTST) approximation\cite{voth1989rigorous} to the reaction rate\cite{predescu2005optimal,  wigner1937calculation, wigner1938transition} 
\begin{align}  \label{qtst}
k(\beta) &\simeq k_{\rm QTST} (\beta, q^\ddagger_{\rm o}) = \min_{q^\ddagger} \left[ k_{\rm QTST} (\beta, q^\ddagger) \right],
\end{align}
with a dividing surface $q_{\rm o}^\ddagger$ that minimizes dynamical recrossing effects. Using the path-integral representation of the QTST rate in the case of a single-surface system
\begin{align}
k_{\rm QTST} (\beta, q_{\rm o}^\ddagger) Q_r
= &  \lim_{n \to \infty } \left(\frac{n}{2\pi \hbar}\right)^n \hspace{-2pt}
\int \hspace{-2pt} d\textbf{p}_0 d\textbf{q}_0
\hspace{2pt}  e^{-\beta H_n^{\rm iso}(\textbf{p}, \textbf{q})}    \nonumber \\
\hspace{2pt}  \times & \hspace{2pt}  \delta({\bar{q}_0- q_{\rm o}^\ddagger}) \hspace{2pt}
 \bar{v}_0 \hspace{2pt} h( \bar{v}_0),
\end{align}
together with Euler's theorem for homogeneous functions,\cite{herman1982path} we derive a virial-like expression for the stationary energy-temperature relation, which can be conveniently evaluated using path-integral Monte Carlo sampling methods
\begin{align}  \label{virial}
E_{\rm st} &= \frac{1}{\beta_{\rm st}} +  \left\langle 
\frac1{2n} \sum_\alpha \left( q_\alpha - \bar{q}_0 \right) \frac{\partial  V(q_\alpha)}{\partial q_\alpha} 
+ \frac1n \sum_\alpha V(q_\alpha) \right\rangle_{q_{\rm o}^\ddagger} 
\end{align}
Here, $\langle \mathcal{Q} \rangle_{q_{\rm o}^\ddagger} $ is a constrained ensemble average defined by
\begin{align} 
\langle \mathcal{Q} \rangle_{q_{\rm o}^\ddagger} &=
\frac{
 \int d\textbf{q}_0 \hspace{2pt} \rho_c(\textbf{q}_0) \hspace{2pt}  \mathcal{Q}(\textbf{q}_0)}
{
 \int d\textbf{q}_0 \hspace{2pt} \rho_c(\textbf{q}_0)},
\end{align}
with
\begin{align}
\rho_c(\textbf{q}_0) =  \delta(\bar{q}_0 - q_{\rm o}^\ddagger )  \hspace{2pt}  e^{-\beta U_{\text{spr}}(\textbf{q}_0)}
\hspace{2pt} e^{-\beta_n \sum_\alpha V(q_\alpha)}.
\end{align}
At high temperatures, Eq.~(\ref{virial}) approaches the classical limit for the stationary energy-temperature relation
\begin{align} \label{cllimit}
\lim_{\beta_{\rm st} \to 0} E_{\rm st} = \frac{1}{\beta_{\rm st}} + V(q_{\rm o}^\ddagger),
\end{align}
with $q_{\rm o}^\ddagger$ approaching the barrier top. This classical limit can be also obtained upon the substitution of the classical transition state theory rate,
\begin{align}
\left[ k Q_r \right]^{\rm CTST} = \frac{1}{2 \pi \beta \hbar} e^{ -\beta V(q_{\rm o}^\ddagger) }, 
\end{align}
into Eq.~(\ref{statbeta}), yielding
\begin{align}
 E_{\rm st} & = -  \left. \frac{\partial \log \left[  \hspace{2pt} k^{\rm CTST}(\beta) Q_r(\beta)  \hspace{2pt} \right] }{\partial \beta} \right|_{\beta_{\rm st}} = \frac{1}{\beta_{\rm st}} + V(q_{\rm o}^\ddagger).
\end{align}
\end{appendix}

\bibliography{ref.bib}

\begin{thebibliography}{68}%
\makeatletter
\providecommand \@ifxundefined [1]{%
 \@ifx{#1\undefined}
}%
\providecommand \@ifnum [1]{%
 \ifnum #1\expandafter \@firstoftwo
 \else \expandafter \@secondoftwo
 \fi
}%
\providecommand \@ifx [1]{%
 \ifx #1\expandafter \@firstoftwo
 \else \expandafter \@secondoftwo
 \fi
}%
\providecommand \natexlab [1]{#1}%
\providecommand \enquote  [1]{``#1''}%
\providecommand \bibnamefont  [1]{#1}%
\providecommand \bibfnamefont [1]{#1}%
\providecommand \citenamefont [1]{#1}%
\providecommand \href@noop [0]{\@secondoftwo}%
\providecommand \href [0]{\begingroup \@sanitize@url \@href}%
\providecommand \@href[1]{\@@startlink{#1}\@@href}%
\providecommand \@@href[1]{\endgroup#1\@@endlink}%
\providecommand \@sanitize@url [0]{\catcode `\\12\catcode `\$12\catcode
  `\&12\catcode `\#12\catcode `\^12\catcode `\_12\catcode `\%12\relax}%
\providecommand \@@startlink[1]{}%
\providecommand \@@endlink[0]{}%
\providecommand \url  [0]{\begingroup\@sanitize@url \@url }%
\providecommand \@url [1]{\endgroup\@href {#1}{\urlprefix }}%
\providecommand \urlprefix  [0]{URL }%
\providecommand \Eprint [0]{\href }%
\providecommand \doibase [0]{http://dx.doi.org/}%
\providecommand \selectlanguage [0]{\@gobble}%
\providecommand \bibinfo  [0]{\@secondoftwo}%
\providecommand \bibfield  [0]{\@secondoftwo}%
\providecommand \translation [1]{[#1]}%
\providecommand \BibitemOpen [0]{}%
\providecommand \bibitemStop [0]{}%
\providecommand \bibitemNoStop [0]{.\EOS\space}%
\providecommand \EOS [0]{\spacefactor3000\relax}%
\providecommand \BibitemShut  [1]{\csname bibitem#1\endcsname}%
\let\auto@bib@innerbib\@empty
\bibitem [{\citenamefont {Craig}\ and\ \citenamefont
  {Manolopoulos}(2004)}]{craig2004quantum}%
  \BibitemOpen
  \bibfield  {author} {\bibinfo {author} {\bibfnamefont {I.~R.}\ \bibnamefont
  {Craig}}\ and\ \bibinfo {author} {\bibfnamefont {D.~E.}\ \bibnamefont
  {Manolopoulos}},\ }\href@noop {} {\bibfield  {journal} {\bibinfo  {journal}
  {J. Chem. Phys.}\ }\textbf {\bibinfo {volume} {121}},\ \bibinfo {pages}
  {3368--3373} (\bibinfo {year} {2004})}\BibitemShut {NoStop}%
\bibitem [{\citenamefont {Habershon}\ \emph {et~al.}(2013)\citenamefont
  {Habershon}, \citenamefont {Manolopoulos}, \citenamefont {Markland},\ and\
  \citenamefont {Miller~III}}]{habershon2013ring}%
  \BibitemOpen
  \bibfield  {author} {\bibinfo {author} {\bibfnamefont {S.}~\bibnamefont
  {Habershon}}, \bibinfo {author} {\bibfnamefont {D.~E.}\ \bibnamefont
  {Manolopoulos}}, \bibinfo {author} {\bibfnamefont {T.~E.}\ \bibnamefont
  {Markland}}, \ and\ \bibinfo {author} {\bibfnamefont {T.~F.}\ \bibnamefont
  {Miller~III}},\ }\href@noop {} {\bibfield  {journal} {\bibinfo  {journal}
  {Annu. Rev. Phys. Chem.}\ }\textbf {\bibinfo {volume} {64}},\ \bibinfo
  {pages} {387--413} (\bibinfo {year} {2013})}\BibitemShut {NoStop}%
\bibitem [{\citenamefont {Craig}\ and\ \citenamefont
  {Manolopoulos}(2005{\natexlab{a}})}]{craig2005chemical}%
  \BibitemOpen
  \bibfield  {author} {\bibinfo {author} {\bibfnamefont {I.~R.}\ \bibnamefont
  {Craig}}\ and\ \bibinfo {author} {\bibfnamefont {D.~E.}\ \bibnamefont
  {Manolopoulos}},\ }\href@noop {} {\bibfield  {journal} {\bibinfo  {journal}
  {J. Chem. Phys.}\ }\textbf {\bibinfo {volume} {122}},\ \bibinfo {pages}
  {084106} (\bibinfo {year} {2005}{\natexlab{a}})}\BibitemShut {NoStop}%
\bibitem [{\citenamefont {Craig}\ and\ \citenamefont
  {Manolopoulos}(2005{\natexlab{b}})}]{craig2005refined}%
  \BibitemOpen
  \bibfield  {author} {\bibinfo {author} {\bibfnamefont {I.~R.}\ \bibnamefont
  {Craig}}\ and\ \bibinfo {author} {\bibfnamefont {D.~E.}\ \bibnamefont
  {Manolopoulos}},\ }\href@noop {} {\bibfield  {journal} {\bibinfo  {journal}
  {J. Chem. Phys.}\ }\textbf {\bibinfo {volume} {123}},\ \bibinfo {pages}
  {034102} (\bibinfo {year} {2005}{\natexlab{b}})}\BibitemShut {NoStop}%
\bibitem [{\citenamefont {Habershon}, \citenamefont {Fanourgakis},\ and\
  \citenamefont {Manolopoulos}(2008)}]{habershon2008comparison}%
  \BibitemOpen
  \bibfield  {author} {\bibinfo {author} {\bibfnamefont {S.}~\bibnamefont
  {Habershon}}, \bibinfo {author} {\bibfnamefont {G.~S.}\ \bibnamefont
  {Fanourgakis}}, \ and\ \bibinfo {author} {\bibfnamefont {D.~E.}\ \bibnamefont
  {Manolopoulos}},\ }\href@noop {} {\bibfield  {journal} {\bibinfo  {journal}
  {J. Chem. Phys.}\ }\textbf {\bibinfo {volume} {129}},\ \bibinfo {pages}
  {074501} (\bibinfo {year} {2008})}\BibitemShut {NoStop}%
\bibitem [{\citenamefont {Rossi}, \citenamefont {Ceriotti},\ and\ \citenamefont
  {Manolopoulos}(2014)}]{rossi2014remove}%
  \BibitemOpen
  \bibfield  {author} {\bibinfo {author} {\bibfnamefont {M.}~\bibnamefont
  {Rossi}}, \bibinfo {author} {\bibfnamefont {M.}~\bibnamefont {Ceriotti}}, \
  and\ \bibinfo {author} {\bibfnamefont {D.~E.}\ \bibnamefont {Manolopoulos}},\
  }\href@noop {} {\bibfield  {journal} {\bibinfo  {journal} {J. Chem. Phys.}\
  }\textbf {\bibinfo {volume} {140}},\ \bibinfo {pages} {234116} (\bibinfo
  {year} {2014})}\BibitemShut {NoStop}%
\bibitem [{\citenamefont {Miller~III}\ and\ \citenamefont
  {Manolopoulos}(2005{\natexlab{a}})}]{miller2005quantum}%
  \BibitemOpen
  \bibfield  {author} {\bibinfo {author} {\bibfnamefont {T.~F.}\ \bibnamefont
  {Miller~III}}\ and\ \bibinfo {author} {\bibfnamefont {D.~E.}\ \bibnamefont
  {Manolopoulos}},\ }\href@noop {} {\bibfield  {journal} {\bibinfo  {journal}
  {J. Chem. Phys.}\ }\textbf {\bibinfo {volume} {122}},\ \bibinfo {pages}
  {184503} (\bibinfo {year} {2005}{\natexlab{a}})}\BibitemShut {NoStop}%
\bibitem [{\citenamefont {Miller~III}\ and\ \citenamefont
  {Manolopoulos}(2005{\natexlab{b}})}]{miller2005quantumwater}%
  \BibitemOpen
  \bibfield  {author} {\bibinfo {author} {\bibfnamefont {T.~F.}\ \bibnamefont
  {Miller~III}}\ and\ \bibinfo {author} {\bibfnamefont {D.~E.}\ \bibnamefont
  {Manolopoulos}},\ }\href@noop {} {\bibfield  {journal} {\bibinfo  {journal}
  {J. Chem. Phys.}\ }\textbf {\bibinfo {volume} {123}},\ \bibinfo {pages}
  {154504} (\bibinfo {year} {2005}{\natexlab{b}})}\BibitemShut {NoStop}%
\bibitem [{\citenamefont {Jiang}\ \emph {et~al.}(2019)\citenamefont {Jiang},
  \citenamefont {Kammler}, \citenamefont {Ding}, \citenamefont {Dorenkamp},
  \citenamefont {Manby}, \citenamefont {Wodtke}, \citenamefont {Miller},
  \citenamefont {Kandratsenka},\ and\ \citenamefont
  {B{\"u}nermann}}]{jiang2019imaging}%
  \BibitemOpen
  \bibfield  {author} {\bibinfo {author} {\bibfnamefont {H.}~\bibnamefont
  {Jiang}}, \bibinfo {author} {\bibfnamefont {M.}~\bibnamefont {Kammler}},
  \bibinfo {author} {\bibfnamefont {F.}~\bibnamefont {Ding}}, \bibinfo {author}
  {\bibfnamefont {Y.}~\bibnamefont {Dorenkamp}}, \bibinfo {author}
  {\bibfnamefont {F.~R.}\ \bibnamefont {Manby}}, \bibinfo {author}
  {\bibfnamefont {A.~M.}\ \bibnamefont {Wodtke}}, \bibinfo {author}
  {\bibfnamefont {T.~F.}\ \bibnamefont {Miller}}, \bibinfo {author}
  {\bibfnamefont {A.}~\bibnamefont {Kandratsenka}}, \ and\ \bibinfo {author}
  {\bibfnamefont {O.}~\bibnamefont {B{\"u}nermann}},\ }\href@noop {} {\bibfield
   {journal} {\bibinfo  {journal} {Science}\ }\textbf {\bibinfo {volume}
  {364}},\ \bibinfo {pages} {379--382} (\bibinfo {year} {2019})}\BibitemShut
  {NoStop}%
\bibitem [{\citenamefont {Wilkins}\ \emph {et~al.}(2017)\citenamefont
  {Wilkins}, \citenamefont {Manolopoulos}, \citenamefont {Pipolo},
  \citenamefont {Laage},\ and\ \citenamefont {Hynes}}]{wilkins2017nuclear}%
  \BibitemOpen
  \bibfield  {author} {\bibinfo {author} {\bibfnamefont {D.~M.}\ \bibnamefont
  {Wilkins}}, \bibinfo {author} {\bibfnamefont {D.~E.}\ \bibnamefont
  {Manolopoulos}}, \bibinfo {author} {\bibfnamefont {S.}~\bibnamefont
  {Pipolo}}, \bibinfo {author} {\bibfnamefont {D.}~\bibnamefont {Laage}}, \
  and\ \bibinfo {author} {\bibfnamefont {J.~T.}\ \bibnamefont {Hynes}},\
  }\href@noop {} {\bibfield  {journal} {\bibinfo  {journal} {J. Phys. Chem.
  Lett.}\ }\textbf {\bibinfo {volume} {8}},\ \bibinfo {pages} {2602--2607}
  (\bibinfo {year} {2017})}\BibitemShut {NoStop}%
\bibitem [{\citenamefont {Rossi}, \citenamefont {Ceriotti},\ and\ \citenamefont
  {Manolopoulos}(2016)}]{rossi2016nuclear}%
  \BibitemOpen
  \bibfield  {author} {\bibinfo {author} {\bibfnamefont {M.}~\bibnamefont
  {Rossi}}, \bibinfo {author} {\bibfnamefont {M.}~\bibnamefont {Ceriotti}}, \
  and\ \bibinfo {author} {\bibfnamefont {D.~E.}\ \bibnamefont {Manolopoulos}},\
  }\href@noop {} {\bibfield  {journal} {\bibinfo  {journal} {J. Phys. Chem.
  Lett.}\ }\textbf {\bibinfo {volume} {7}},\ \bibinfo {pages} {3001--3007}
  (\bibinfo {year} {2016})}\BibitemShut {NoStop}%
\bibitem [{\citenamefont {Ceriotti}\ \emph {et~al.}(2013)\citenamefont
  {Ceriotti}, \citenamefont {Cuny}, \citenamefont {Parrinello},\ and\
  \citenamefont {Manolopoulos}}]{ceriotti2013nuclear}%
  \BibitemOpen
  \bibfield  {author} {\bibinfo {author} {\bibfnamefont {M.}~\bibnamefont
  {Ceriotti}}, \bibinfo {author} {\bibfnamefont {J.}~\bibnamefont {Cuny}},
  \bibinfo {author} {\bibfnamefont {M.}~\bibnamefont {Parrinello}}, \ and\
  \bibinfo {author} {\bibfnamefont {D.~E.}\ \bibnamefont {Manolopoulos}},\
  }\href@noop {} {\bibfield  {journal} {\bibinfo  {journal} {Proc. Natl. Acad.
  Sci. U.S.A.}\ }\textbf {\bibinfo {volume} {110}},\ \bibinfo {pages}
  {15591--15596} (\bibinfo {year} {2013})}\BibitemShut {NoStop}%
\bibitem [{\citenamefont {Marsalek}\ and\ \citenamefont
  {Markland}(2016)}]{marsalek2016ab}%
  \BibitemOpen
  \bibfield  {author} {\bibinfo {author} {\bibfnamefont {O.}~\bibnamefont
  {Marsalek}}\ and\ \bibinfo {author} {\bibfnamefont {T.~E.}\ \bibnamefont
  {Markland}},\ }\href@noop {} {\bibfield  {journal} {\bibinfo  {journal} {J.
  Chem. Phys.}\ }\textbf {\bibinfo {volume} {144}},\ \bibinfo {pages} {054112}
  (\bibinfo {year} {2016})}\BibitemShut {NoStop}%
\bibitem [{\citenamefont {Menzeleev}, \citenamefont {Ananth},\ and\
  \citenamefont {Miller~III}(2011)}]{menzeleev2011direct}%
  \BibitemOpen
  \bibfield  {author} {\bibinfo {author} {\bibfnamefont {A.~R.}\ \bibnamefont
  {Menzeleev}}, \bibinfo {author} {\bibfnamefont {N.}~\bibnamefont {Ananth}}, \
  and\ \bibinfo {author} {\bibfnamefont {T.~F.}\ \bibnamefont {Miller~III}},\
  }\href@noop {} {\bibfield  {journal} {\bibinfo  {journal} {J. Chem. Phys.}\
  }\textbf {\bibinfo {volume} {135}},\ \bibinfo {pages} {074106} (\bibinfo
  {year} {2011})}\BibitemShut {NoStop}%
\bibitem [{\citenamefont {Menzeleev}\ and\ \citenamefont
  {Miller~III}(2010)}]{menzeleev2010ring}%
  \BibitemOpen
  \bibfield  {author} {\bibinfo {author} {\bibfnamefont {A.~R.}\ \bibnamefont
  {Menzeleev}}\ and\ \bibinfo {author} {\bibfnamefont {T.~F.}\ \bibnamefont
  {Miller~III}},\ }\href@noop {} {\bibfield  {journal} {\bibinfo  {journal} {J.
  Chem. Phys.}\ }\textbf {\bibinfo {volume} {132}},\ \bibinfo {pages} {034106}
  (\bibinfo {year} {2010})}\BibitemShut {NoStop}%
\bibitem [{\citenamefont {Kretchmer}\ and\ \citenamefont
  {Miller~III}(2013)}]{kretchmer2013direct}%
  \BibitemOpen
  \bibfield  {author} {\bibinfo {author} {\bibfnamefont {J.~S.}\ \bibnamefont
  {Kretchmer}}\ and\ \bibinfo {author} {\bibfnamefont {T.~F.}\ \bibnamefont
  {Miller~III}},\ }\href@noop {} {\bibfield  {journal} {\bibinfo  {journal} {J.
  Chem. Phys.}\ }\textbf {\bibinfo {volume} {138}},\ \bibinfo {pages} {04B602}
  (\bibinfo {year} {2013})}\BibitemShut {NoStop}%
\bibitem [{\citenamefont {Kretchmer}\ and\ \citenamefont
  {Miller~III}(2015)}]{kretchmer2015tipping}%
  \BibitemOpen
  \bibfield  {author} {\bibinfo {author} {\bibfnamefont {J.~S.}\ \bibnamefont
  {Kretchmer}}\ and\ \bibinfo {author} {\bibfnamefont {T.~F.}\ \bibnamefont
  {Miller~III}},\ }\href@noop {} {\bibfield  {journal} {\bibinfo  {journal}
  {Inorg. Chem.}\ }\textbf {\bibinfo {volume} {55}},\ \bibinfo {pages}
  {1022--1031} (\bibinfo {year} {2015})}\BibitemShut {NoStop}%
\bibitem [{\citenamefont {Suleimanov}, \citenamefont {Aoiz},\ and\
  \citenamefont {Guo}(2016)}]{suleimanov2016chemical}%
  \BibitemOpen
  \bibfield  {author} {\bibinfo {author} {\bibfnamefont {Y.~V.}\ \bibnamefont
  {Suleimanov}}, \bibinfo {author} {\bibfnamefont {F.~J.}\ \bibnamefont
  {Aoiz}}, \ and\ \bibinfo {author} {\bibfnamefont {H.}~\bibnamefont {Guo}},\
  }\href@noop {} {\bibfield  {journal} {\bibinfo  {journal} {J. Phys. Chem. A}\
  }\textbf {\bibinfo {volume} {120}},\ \bibinfo {pages} {8488--8502} (\bibinfo
  {year} {2016})}\BibitemShut {NoStop}%
\bibitem [{\citenamefont {Collepardo-Guevara}, \citenamefont {Suleimanov},\
  and\ \citenamefont {Manolopoulos}(2009)}]{collepardo2009bimolecular}%
  \BibitemOpen
  \bibfield  {author} {\bibinfo {author} {\bibfnamefont {R.}~\bibnamefont
  {Collepardo-Guevara}}, \bibinfo {author} {\bibfnamefont {Y.~V.}\ \bibnamefont
  {Suleimanov}}, \ and\ \bibinfo {author} {\bibfnamefont {D.~E.}\ \bibnamefont
  {Manolopoulos}},\ }\href@noop {} {\bibfield  {journal} {\bibinfo  {journal}
  {J. Chem. Phys.}\ }\textbf {\bibinfo {volume} {130}},\ \bibinfo {pages}
  {174713} (\bibinfo {year} {2009})}\BibitemShut {NoStop}%
\bibitem [{\citenamefont {Perez~de Tudela}\ \emph {et~al.}(2012)\citenamefont
  {Perez~de Tudela}, \citenamefont {Aoiz}, \citenamefont {Suleimanov},\ and\
  \citenamefont {Manolopoulos}}]{perez2012chemical}%
  \BibitemOpen
  \bibfield  {author} {\bibinfo {author} {\bibfnamefont {R.}~\bibnamefont
  {Perez~de Tudela}}, \bibinfo {author} {\bibfnamefont {F.}~\bibnamefont
  {Aoiz}}, \bibinfo {author} {\bibfnamefont {Y.~V.}\ \bibnamefont
  {Suleimanov}}, \ and\ \bibinfo {author} {\bibfnamefont {D.~E.}\ \bibnamefont
  {Manolopoulos}},\ }\href@noop {} {\bibfield  {journal} {\bibinfo  {journal}
  {J. Phys. Chem. Lett.}\ }\textbf {\bibinfo {volume} {3}},\ \bibinfo {pages}
  {493--497} (\bibinfo {year} {2012})}\BibitemShut {NoStop}%
\bibitem [{\citenamefont {Li}\ \emph {et~al.}(2012)\citenamefont {Li},
  \citenamefont {Suleimanov}, \citenamefont {Yang}, \citenamefont {Green},\
  and\ \citenamefont {Guo}}]{li2012ring}%
  \BibitemOpen
  \bibfield  {author} {\bibinfo {author} {\bibfnamefont {Y.}~\bibnamefont
  {Li}}, \bibinfo {author} {\bibfnamefont {Y.~V.}\ \bibnamefont {Suleimanov}},
  \bibinfo {author} {\bibfnamefont {M.}~\bibnamefont {Yang}}, \bibinfo {author}
  {\bibfnamefont {W.~H.}\ \bibnamefont {Green}}, \ and\ \bibinfo {author}
  {\bibfnamefont {H.}~\bibnamefont {Guo}},\ }\href@noop {} {\bibfield
  {journal} {\bibinfo  {journal} {J. Phys. Chem. Lett.}\ }\textbf {\bibinfo
  {volume} {4}},\ \bibinfo {pages} {48--52} (\bibinfo {year}
  {2012})}\BibitemShut {NoStop}%
\bibitem [{\citenamefont {Li}\ \emph {et~al.}(2013)\citenamefont {Li},
  \citenamefont {Suleimanov}, \citenamefont {Li}, \citenamefont {Green},\ and\
  \citenamefont {Guo}}]{li2013rate}%
  \BibitemOpen
  \bibfield  {author} {\bibinfo {author} {\bibfnamefont {Y.}~\bibnamefont
  {Li}}, \bibinfo {author} {\bibfnamefont {Y.~V.}\ \bibnamefont {Suleimanov}},
  \bibinfo {author} {\bibfnamefont {J.}~\bibnamefont {Li}}, \bibinfo {author}
  {\bibfnamefont {W.~H.}\ \bibnamefont {Green}}, \ and\ \bibinfo {author}
  {\bibfnamefont {H.}~\bibnamefont {Guo}},\ }\href@noop {} {\bibfield
  {journal} {\bibinfo  {journal} {J. Chem. Phys.}\ }\textbf {\bibinfo {volume}
  {138}},\ \bibinfo {pages} {094307} (\bibinfo {year} {2013})}\BibitemShut
  {NoStop}%
\bibitem [{\citenamefont {Suleimanov}(2012)}]{suleimanov2012surface}%
  \BibitemOpen
  \bibfield  {author} {\bibinfo {author} {\bibfnamefont {Y.~V.}\ \bibnamefont
  {Suleimanov}},\ }\href@noop {} {\bibfield  {journal} {\bibinfo  {journal} {J.
  Phys. Chem. C}\ }\textbf {\bibinfo {volume} {116}},\ \bibinfo {pages}
  {11141--11153} (\bibinfo {year} {2012})}\BibitemShut {NoStop}%
\bibitem [{\citenamefont {Boekelheide}, \citenamefont {Salom{\'o}n-Ferrer},\
  and\ \citenamefont {Miller}(2011)}]{boekelheide2011dynamics}%
  \BibitemOpen
  \bibfield  {author} {\bibinfo {author} {\bibfnamefont {N.}~\bibnamefont
  {Boekelheide}}, \bibinfo {author} {\bibfnamefont {R.}~\bibnamefont
  {Salom{\'o}n-Ferrer}}, \ and\ \bibinfo {author} {\bibfnamefont {T.~F.}\
  \bibnamefont {Miller}},\ }\href@noop {} {\bibfield  {journal} {\bibinfo
  {journal} {Proc. Natl. Acad. Sci. U.S.A.}\ }\textbf {\bibinfo {volume}
  {108}},\ \bibinfo {pages} {16159--16163} (\bibinfo {year}
  {2011})}\BibitemShut {NoStop}%
\bibitem [{\citenamefont {Miller~III}(2008)}]{miller2008isomorphic}%
  \BibitemOpen
  \bibfield  {author} {\bibinfo {author} {\bibfnamefont {T.~F.}\ \bibnamefont
  {Miller~III}},\ }\href@noop {} {\bibfield  {journal} {\bibinfo  {journal} {J.
  Chem. Phys.}\ }\textbf {\bibinfo {volume} {129}},\ \bibinfo {pages} {194502}
  (\bibinfo {year} {2008})}\BibitemShut {NoStop}%
\bibitem [{\citenamefont {Kreis}\ \emph {et~al.}(2017)\citenamefont {Kreis},
  \citenamefont {Kremer}, \citenamefont {Potestio},\ and\ \citenamefont
  {Tuckerman}}]{kreis2017classical}%
  \BibitemOpen
  \bibfield  {author} {\bibinfo {author} {\bibfnamefont {K.}~\bibnamefont
  {Kreis}}, \bibinfo {author} {\bibfnamefont {K.}~\bibnamefont {Kremer}},
  \bibinfo {author} {\bibfnamefont {R.}~\bibnamefont {Potestio}}, \ and\
  \bibinfo {author} {\bibfnamefont {M.~E.}\ \bibnamefont {Tuckerman}},\
  }\href@noop {} {\bibfield  {journal} {\bibinfo  {journal} {J. Chem. Phys.}\
  }\textbf {\bibinfo {volume} {147}},\ \bibinfo {pages} {244104} (\bibinfo
  {year} {2017})}\BibitemShut {NoStop}%
\bibitem [{\citenamefont {Hele}()}]{hele2005an}%
  \BibitemOpen
  \bibfield  {author} {\bibinfo {author} {\bibfnamefont {T.~J.~H.}\
  \bibnamefont {Hele}},\ }\href@noop {} {}\bibinfo {note} {Master's Thesis,
  Exeter College, Oxford University, 2011}\BibitemShut {NoStop}%
\bibitem [{\citenamefont {Duke}\ and\ \citenamefont
  {Ananth}(2017)}]{duke2017mean}%
  \BibitemOpen
  \bibfield  {author} {\bibinfo {author} {\bibfnamefont {J.~R.}\ \bibnamefont
  {Duke}}\ and\ \bibinfo {author} {\bibfnamefont {N.}~\bibnamefont {Ananth}},\
  }\href@noop {} {\bibfield  {journal} {\bibinfo  {journal} {Faraday Discuss.}\
  }\textbf {\bibinfo {volume} {195}},\ \bibinfo {pages} {253--268} (\bibinfo
  {year} {2017})}\BibitemShut {NoStop}%
\bibitem [{\citenamefont {Menzeleev}, \citenamefont {Bell},\ and\ \citenamefont
  {Miller~III}(2014)}]{menzeleev2014kinetically}%
  \BibitemOpen
  \bibfield  {author} {\bibinfo {author} {\bibfnamefont {A.~R.}\ \bibnamefont
  {Menzeleev}}, \bibinfo {author} {\bibfnamefont {F.}~\bibnamefont {Bell}}, \
  and\ \bibinfo {author} {\bibfnamefont {T.~F.}\ \bibnamefont {Miller~III}},\
  }\href@noop {} {\bibfield  {journal} {\bibinfo  {journal} {J. Chem. Phys.}\
  }\textbf {\bibinfo {volume} {140}},\ \bibinfo {pages} {064103} (\bibinfo
  {year} {2014})}\BibitemShut {NoStop}%
\bibitem [{\citenamefont {Kretchmer}\ and\ \citenamefont
  {Miller~III}(2017)}]{kretchmer2017kinetically}%
  \BibitemOpen
  \bibfield  {author} {\bibinfo {author} {\bibfnamefont {J.~S.}\ \bibnamefont
  {Kretchmer}}\ and\ \bibinfo {author} {\bibfnamefont {T.~F.}\ \bibnamefont
  {Miller~III}},\ }\href@noop {} {\bibfield  {journal} {\bibinfo  {journal}
  {Faraday Discuss.}\ }\textbf {\bibinfo {volume} {195}},\ \bibinfo {pages}
  {191--214} (\bibinfo {year} {2017})}\BibitemShut {NoStop}%
\bibitem [{\citenamefont {Kretchmer}\ \emph {et~al.}(2018)\citenamefont
  {Kretchmer}, \citenamefont {Boekelheide}, \citenamefont {Warren},
  \citenamefont {Winkler}, \citenamefont {Gray},\ and\ \citenamefont
  {Miller~III}}]{kretchmer2018fluctuating}%
  \BibitemOpen
  \bibfield  {author} {\bibinfo {author} {\bibfnamefont {J.~S.}\ \bibnamefont
  {Kretchmer}}, \bibinfo {author} {\bibfnamefont {N.}~\bibnamefont
  {Boekelheide}}, \bibinfo {author} {\bibfnamefont {J.~J.}\ \bibnamefont
  {Warren}}, \bibinfo {author} {\bibfnamefont {J.~R.}\ \bibnamefont {Winkler}},
  \bibinfo {author} {\bibfnamefont {H.~B.}\ \bibnamefont {Gray}}, \ and\
  \bibinfo {author} {\bibfnamefont {T.~F.}\ \bibnamefont {Miller~III}},\
  }\href@noop {} {\bibfield  {journal} {\bibinfo  {journal} {Proc. Natl. Acad.
  Sci. U.S.A.}\ }\textbf {\bibinfo {volume} {115}},\ \bibinfo {pages}
  {6129--6134} (\bibinfo {year} {2018})}\BibitemShut {NoStop}%
\bibitem [{\citenamefont {Ananth}(2013)}]{ananth2013mapping}%
  \BibitemOpen
  \bibfield  {author} {\bibinfo {author} {\bibfnamefont {N.}~\bibnamefont
  {Ananth}},\ }\href@noop {} {\bibfield  {journal} {\bibinfo  {journal} {J.
  Chem. Phys.}\ }\textbf {\bibinfo {volume} {139}},\ \bibinfo {pages} {124102}
  (\bibinfo {year} {2013})}\BibitemShut {NoStop}%
\bibitem [{\citenamefont {Pierre}\ \emph {et~al.}(2017)\citenamefont {Pierre},
  \citenamefont {Duke}, \citenamefont {Hele},\ and\ \citenamefont
  {Ananth}}]{pierre2017mapping}%
  \BibitemOpen
  \bibfield  {author} {\bibinfo {author} {\bibfnamefont {S.}~\bibnamefont
  {Pierre}}, \bibinfo {author} {\bibfnamefont {J.~R.}\ \bibnamefont {Duke}},
  \bibinfo {author} {\bibfnamefont {T.~J.}\ \bibnamefont {Hele}}, \ and\
  \bibinfo {author} {\bibfnamefont {N.}~\bibnamefont {Ananth}},\ }\href@noop {}
  {\bibfield  {journal} {\bibinfo  {journal} {J. Chem. Phys.}\ }\textbf
  {\bibinfo {volume} {147}},\ \bibinfo {pages} {234103} (\bibinfo {year}
  {2017})}\BibitemShut {NoStop}%
\bibitem [{\citenamefont {Richardson}\ and\ \citenamefont
  {Thoss}(2013)}]{richardson2013communication}%
  \BibitemOpen
  \bibfield  {author} {\bibinfo {author} {\bibfnamefont {J.~O.}\ \bibnamefont
  {Richardson}}\ and\ \bibinfo {author} {\bibfnamefont {M.}~\bibnamefont
  {Thoss}},\ }\href@noop {} {\bibfield  {journal} {\bibinfo  {journal} {J.
  Chem. Phys.}\ }\textbf {\bibinfo {volume} {139}},\ \bibinfo {pages} {031102}
  (\bibinfo {year} {2013})}\BibitemShut {NoStop}%
\bibitem [{\citenamefont {Chowdhury}\ and\ \citenamefont
  {Huo}(2017)}]{chowdhury2017coherent}%
  \BibitemOpen
  \bibfield  {author} {\bibinfo {author} {\bibfnamefont {S.~N.}\ \bibnamefont
  {Chowdhury}}\ and\ \bibinfo {author} {\bibfnamefont {P.}~\bibnamefont
  {Huo}},\ }\href@noop {} {\bibfield  {journal} {\bibinfo  {journal} {J. Chem.
  Phys.}\ }\textbf {\bibinfo {volume} {147}},\ \bibinfo {pages} {214109}
  (\bibinfo {year} {2017})}\BibitemShut {NoStop}%
\bibitem [{\citenamefont {Tao}, \citenamefont {Shushkov},\ and\ \citenamefont
  {Miller~III}(2018)}]{tao2018path}%
  \BibitemOpen
  \bibfield  {author} {\bibinfo {author} {\bibfnamefont {X.}~\bibnamefont
  {Tao}}, \bibinfo {author} {\bibfnamefont {P.}~\bibnamefont {Shushkov}}, \
  and\ \bibinfo {author} {\bibfnamefont {T.~F.}\ \bibnamefont {Miller~III}},\
  }\href@noop {} {\bibfield  {journal} {\bibinfo  {journal} {J. Chem. Phys.}\
  }\textbf {\bibinfo {volume} {148}},\ \bibinfo {pages} {102327} (\bibinfo
  {year} {2018})}\BibitemShut {NoStop}%
\bibitem [{\citenamefont {Welsch}\ \emph {et~al.}(2016)\citenamefont {Welsch},
  \citenamefont {Song}, \citenamefont {Shi}, \citenamefont {Althorpe},\ and\
  \citenamefont {Miller~III}}]{welsch2016non}%
  \BibitemOpen
  \bibfield  {author} {\bibinfo {author} {\bibfnamefont {R.}~\bibnamefont
  {Welsch}}, \bibinfo {author} {\bibfnamefont {K.}~\bibnamefont {Song}},
  \bibinfo {author} {\bibfnamefont {Q.}~\bibnamefont {Shi}}, \bibinfo {author}
  {\bibfnamefont {S.~C.}\ \bibnamefont {Althorpe}}, \ and\ \bibinfo {author}
  {\bibfnamefont {T.~F.}\ \bibnamefont {Miller~III}},\ }\href@noop {}
  {\bibfield  {journal} {\bibinfo  {journal} {J. Chem. Phys.}\ }\textbf
  {\bibinfo {volume} {145}},\ \bibinfo {pages} {204118} (\bibinfo {year}
  {2016})}\BibitemShut {NoStop}%
\bibitem [{\citenamefont {Duke}\ and\ \citenamefont
  {Ananth}(2015)}]{duke2015simulating}%
  \BibitemOpen
  \bibfield  {author} {\bibinfo {author} {\bibfnamefont {J.~R.}\ \bibnamefont
  {Duke}}\ and\ \bibinfo {author} {\bibfnamefont {N.}~\bibnamefont {Ananth}},\
  }\href@noop {} {\bibfield  {journal} {\bibinfo  {journal} {J. Phys. Chem.
  Lett.}\ }\textbf {\bibinfo {volume} {6}},\ \bibinfo {pages} {4219--4223}
  (\bibinfo {year} {2015})}\BibitemShut {NoStop}%
\bibitem [{\citenamefont {Shakib}\ and\ \citenamefont
  {Huo}(2017)}]{shakib2017ring}%
  \BibitemOpen
  \bibfield  {author} {\bibinfo {author} {\bibfnamefont {F.~A.}\ \bibnamefont
  {Shakib}}\ and\ \bibinfo {author} {\bibfnamefont {P.}~\bibnamefont {Huo}},\
  }\href@noop {} {\bibfield  {journal} {\bibinfo  {journal} {J. Phys. Chem.
  Lett.}\ }\textbf {\bibinfo {volume} {8}},\ \bibinfo {pages} {3073--3080}
  (\bibinfo {year} {2017})}\BibitemShut {NoStop}%
\bibitem [{\citenamefont {Feynman}\ and\ \citenamefont
  {Hibbs}(1965)}]{feynman1965quantum}%
  \BibitemOpen
  \bibfield  {author} {\bibinfo {author} {\bibfnamefont {R.~P.}\ \bibnamefont
  {Feynman}}\ and\ \bibinfo {author} {\bibfnamefont {A.~R.}\ \bibnamefont
  {Hibbs}},\ }\href@noop {} {\emph {\bibinfo {title} {Quantum mechanics and
  path integrals}}}\ (\bibinfo  {publisher} {McGraw-Hill},\ \bibinfo {year}
  {1965})\BibitemShut {NoStop}%
\bibitem [{\citenamefont {Chandler}\ and\ \citenamefont
  {Wolynes}(1981)}]{chandler1981exploiting}%
  \BibitemOpen
  \bibfield  {author} {\bibinfo {author} {\bibfnamefont {D.}~\bibnamefont
  {Chandler}}\ and\ \bibinfo {author} {\bibfnamefont {P.~G.}\ \bibnamefont
  {Wolynes}},\ }\href@noop {} {\bibfield  {journal} {\bibinfo  {journal} {J.
  Chem. Phys.}\ }\textbf {\bibinfo {volume} {74}},\ \bibinfo {pages}
  {4078--4095} (\bibinfo {year} {1981})}\BibitemShut {NoStop}%
\bibitem [{\citenamefont {Parrinello}\ and\ \citenamefont
  {Rahman}(1984)}]{parrinello1984study}%
  \BibitemOpen
  \bibfield  {author} {\bibinfo {author} {\bibfnamefont {M.}~\bibnamefont
  {Parrinello}}\ and\ \bibinfo {author} {\bibfnamefont {A.}~\bibnamefont
  {Rahman}},\ }\href@noop {} {\bibfield  {journal} {\bibinfo  {journal} {J.
  Chem. Phys.}\ }\textbf {\bibinfo {volume} {80}},\ \bibinfo {pages} {860--867}
  (\bibinfo {year} {1984})}\BibitemShut {NoStop}%
\bibitem [{\citenamefont {Miller}(1975)}]{miller1975path}%
  \BibitemOpen
  \bibfield  {author} {\bibinfo {author} {\bibfnamefont {W.~H.}\ \bibnamefont
  {Miller}},\ }\href@noop {} {\bibfield  {journal} {\bibinfo  {journal} {J.
  Chem. Phys.}\ }\textbf {\bibinfo {volume} {63}},\ \bibinfo {pages}
  {1166--1172} (\bibinfo {year} {1975})}\BibitemShut {NoStop}%
\bibitem [{\citenamefont {Press}(2007)}]{press2007numerical}%
  \BibitemOpen
  \bibfield  {author} {\bibinfo {author} {\bibfnamefont {W.~H.}\ \bibnamefont
  {Press}},\ }\href@noop {} {\emph {\bibinfo {title} {Numerical recipes 3rd
  edition: The art of scientific computing}}}\ (\bibinfo  {publisher}
  {Cambridge university press},\ \bibinfo {year} {2007})\BibitemShut {NoStop}%
\bibitem [{\citenamefont {Bryan}(1990)}]{bryan1990maximum}%
  \BibitemOpen
  \bibfield  {author} {\bibinfo {author} {\bibfnamefont {R.}~\bibnamefont
  {Bryan}},\ }\href@noop {} {\bibfield  {journal} {\bibinfo  {journal} {Eur.
  Biophys. J.}\ }\textbf {\bibinfo {volume} {18}},\ \bibinfo {pages} {165--174}
  (\bibinfo {year} {1990})}\BibitemShut {NoStop}%
\bibitem [{\citenamefont {Jarrell}\ and\ \citenamefont
  {Gubernatis}(1996)}]{jarrell1996bayesian}%
  \BibitemOpen
  \bibfield  {author} {\bibinfo {author} {\bibfnamefont {M.}~\bibnamefont
  {Jarrell}}\ and\ \bibinfo {author} {\bibfnamefont {J.~E.}\ \bibnamefont
  {Gubernatis}},\ }\href@noop {} {\bibfield  {journal} {\bibinfo  {journal}
  {Phys. Rep.}\ }\textbf {\bibinfo {volume} {269}},\ \bibinfo {pages}
  {133--195} (\bibinfo {year} {1996})}\BibitemShut {NoStop}%
\bibitem [{\citenamefont {Habershon}, \citenamefont {Braams},\ and\
  \citenamefont {Manolopoulos}(2007)}]{habershon2007quantum}%
  \BibitemOpen
  \bibfield  {author} {\bibinfo {author} {\bibfnamefont {S.}~\bibnamefont
  {Habershon}}, \bibinfo {author} {\bibfnamefont {B.~J.}\ \bibnamefont
  {Braams}}, \ and\ \bibinfo {author} {\bibfnamefont {D.~E.}\ \bibnamefont
  {Manolopoulos}},\ }\href@noop {} {\bibfield  {journal} {\bibinfo  {journal}
  {J. Chem. Phys.}\ }\textbf {\bibinfo {volume} {127}},\ \bibinfo {pages}
  {174108} (\bibinfo {year} {2007})}\BibitemShut {NoStop}%
\bibitem [{\citenamefont {Rabani}, \citenamefont {Krilov},\ and\ \citenamefont
  {Berne}(2000)}]{rabani2000quantum}%
  \BibitemOpen
  \bibfield  {author} {\bibinfo {author} {\bibfnamefont {E.}~\bibnamefont
  {Rabani}}, \bibinfo {author} {\bibfnamefont {G.}~\bibnamefont {Krilov}}, \
  and\ \bibinfo {author} {\bibfnamefont {B.}~\bibnamefont {Berne}},\
  }\href@noop {} {\bibfield  {journal} {\bibinfo  {journal} {J. Chem. Phys.}\
  }\textbf {\bibinfo {volume} {112}},\ \bibinfo {pages} {2605--2614} (\bibinfo
  {year} {2000})}\BibitemShut {NoStop}%
\bibitem [{\citenamefont {Golosov}, \citenamefont {Reichman},\ and\
  \citenamefont {Rabani}(2003)}]{golosov2003analytic}%
  \BibitemOpen
  \bibfield  {author} {\bibinfo {author} {\bibfnamefont {A.~A.}\ \bibnamefont
  {Golosov}}, \bibinfo {author} {\bibfnamefont {D.~R.}\ \bibnamefont
  {Reichman}}, \ and\ \bibinfo {author} {\bibfnamefont {E.}~\bibnamefont
  {Rabani}},\ }\href@noop {} {\bibfield  {journal} {\bibinfo  {journal} {J.
  Chem. Phys.}\ }\textbf {\bibinfo {volume} {118}},\ \bibinfo {pages}
  {457--460} (\bibinfo {year} {2003})}\BibitemShut {NoStop}%
\bibitem [{\citenamefont {Miller}(2006)}]{miller2006applied}%
  \BibitemOpen
  \bibfield  {author} {\bibinfo {author} {\bibfnamefont {P.~D.}\ \bibnamefont
  {Miller}},\ }\href@noop {} {\emph {\bibinfo {title} {Applied asymptotic
  analysis}}}\ (\bibinfo  {publisher} {American Mathematical Society},\
  \bibinfo {year} {2006})\BibitemShut {NoStop}%
\bibitem [{\citenamefont {Miller}\ \emph {et~al.}(2003)\citenamefont {Miller},
  \citenamefont {Zhao}, \citenamefont {Ceotto},\ and\ \citenamefont
  {Yang}}]{miller2003quantum}%
  \BibitemOpen
  \bibfield  {author} {\bibinfo {author} {\bibfnamefont {W.~H.}\ \bibnamefont
  {Miller}}, \bibinfo {author} {\bibfnamefont {Y.}~\bibnamefont {Zhao}},
  \bibinfo {author} {\bibfnamefont {M.}~\bibnamefont {Ceotto}}, \ and\ \bibinfo
  {author} {\bibfnamefont {S.}~\bibnamefont {Yang}},\ }\href@noop {} {\bibfield
   {journal} {\bibinfo  {journal} {J. Chem. Phys.}\ }\textbf {\bibinfo {volume}
  {119}},\ \bibinfo {pages} {1329--1342} (\bibinfo {year} {2003})}\BibitemShut
  {NoStop}%
\bibitem [{\citenamefont {Korol}, \citenamefont {Bou-Rabee},\ and\
  \citenamefont {Miller~III}(2019{\natexlab{a}})}]{korol2019cayley}%
  \BibitemOpen
  \bibfield  {author} {\bibinfo {author} {\bibfnamefont {R.}~\bibnamefont
  {Korol}}, \bibinfo {author} {\bibfnamefont {N.}~\bibnamefont {Bou-Rabee}}, \
  and\ \bibinfo {author} {\bibfnamefont {T.~F.}\ \bibnamefont {Miller~III}},\
  }\href@noop {} {\bibfield  {journal} {\bibinfo  {journal} {J. Chem. Phys.}\
  }\textbf {\bibinfo {volume} {151}},\ \bibinfo {pages} {124103} (\bibinfo
  {year} {2019}{\natexlab{a}})}\BibitemShut {NoStop}%
\bibitem [{\citenamefont {Korol}, \citenamefont {Bou-Rabee},\ and\
  \citenamefont {Miller~III}(2019{\natexlab{b}})}]{korol2019cayley2}%
  \BibitemOpen
  \bibfield  {author} {\bibinfo {author} {\bibfnamefont {R.}~\bibnamefont
  {Korol}}, \bibinfo {author} {\bibfnamefont {N.}~\bibnamefont {Bou-Rabee}}, \
  and\ \bibinfo {author} {\bibfnamefont {T.~F.}\ \bibnamefont {Miller~III}},\
  }\href@noop {} {\bibfield  {journal} {\bibinfo  {journal} {arXiv:1911.00931}\
  } (\bibinfo {year} {2019}{\natexlab{b}})}\BibitemShut {NoStop}%
\bibitem [{\citenamefont {Dierckx}(1995)}]{dierckx1995curve}%
  \BibitemOpen
  \bibfield  {author} {\bibinfo {author} {\bibfnamefont {P.}~\bibnamefont
  {Dierckx}},\ }\href@noop {} {\emph {\bibinfo {title} {Curve and surface
  fitting with splines}}}\ (\bibinfo  {publisher} {Oxford University Press},\
  \bibinfo {year} {1995})\BibitemShut {NoStop}%
\bibitem [{\citenamefont {Jones}\ \emph {et~al.}()\citenamefont {Jones},
  \citenamefont {Oliphant}, \citenamefont {Peterson} \emph {et~al.}}]{scipy}%
  \BibitemOpen
  \bibfield  {author} {\bibinfo {author} {\bibfnamefont {E.}~\bibnamefont
  {Jones}}, \bibinfo {author} {\bibfnamefont {T.}~\bibnamefont {Oliphant}},
  \bibinfo {author} {\bibfnamefont {P.}~\bibnamefont {Peterson}},  \emph
  {et~al.},\ }\href@noop {} {}\bibinfo {note} {SciPy: Open source scientific
  tools for Python, (2001--)}\BibitemShut {NoStop}%
\bibitem [{cod()}]{code}%
  \BibitemOpen
  \href@noop {} {}\bibinfo {note} {See https://github.com/jgreitemann/maxent
  for the available code provided by Jonas Greitemann}\BibitemShut {NoStop}%
\bibitem [{\citenamefont {Richardson}\ and\ \citenamefont
  {Althorpe}(2009)}]{richardson2009ring}%
  \BibitemOpen
  \bibfield  {author} {\bibinfo {author} {\bibfnamefont {J.~O.}\ \bibnamefont
  {Richardson}}\ and\ \bibinfo {author} {\bibfnamefont {S.~C.}\ \bibnamefont
  {Althorpe}},\ }\href@noop {} {\bibfield  {journal} {\bibinfo  {journal} {J.
  Chem. Phys.}\ }\textbf {\bibinfo {volume} {131}},\ \bibinfo {pages} {214106}
  (\bibinfo {year} {2009})}\BibitemShut {NoStop}%
\bibitem [{\citenamefont {Richardson}(2017)}]{richardson2017microcanonical}%
  \BibitemOpen
  \bibfield  {author} {\bibinfo {author} {\bibfnamefont {J.~O.}\ \bibnamefont
  {Richardson}},\ }\href@noop {} {\bibfield  {journal} {\bibinfo  {journal}
  {Faraday Discuss.}\ }\textbf {\bibinfo {volume} {195}},\ \bibinfo {pages}
  {49--67} (\bibinfo {year} {2017})}\BibitemShut {NoStop}%
\bibitem [{\citenamefont {Tao}, \citenamefont {Shushkov},\ and\ \citenamefont
  {Miller~III}(2019)}]{tao2019simple}%
  \BibitemOpen
  \bibfield  {author} {\bibinfo {author} {\bibfnamefont {X.}~\bibnamefont
  {Tao}}, \bibinfo {author} {\bibfnamefont {P.}~\bibnamefont {Shushkov}}, \
  and\ \bibinfo {author} {\bibfnamefont {T.~F.}\ \bibnamefont {Miller~III}},\
  }\href@noop {} {\bibfield  {journal} {\bibinfo  {journal} {J. Phys. Chem. A}\
  }\textbf {\bibinfo {volume} {123}},\ \bibinfo {pages} {3013--3020} (\bibinfo
  {year} {2019})}\BibitemShut {NoStop}%
\bibitem [{\citenamefont {Feit}, \citenamefont {Fleck},\ and\ \citenamefont
  {Steiger}(1982)}]{feit1982solution}%
  \BibitemOpen
  \bibfield  {author} {\bibinfo {author} {\bibfnamefont {M.}~\bibnamefont
  {Feit}}, \bibinfo {author} {\bibfnamefont {J.}~\bibnamefont {Fleck}}, \ and\
  \bibinfo {author} {\bibfnamefont {A.}~\bibnamefont {Steiger}},\ }\href@noop
  {} {\bibfield  {journal} {\bibinfo  {journal} {J. Comput. Phys.}\ }\textbf
  {\bibinfo {volume} {47}},\ \bibinfo {pages} {412--433} (\bibinfo {year}
  {1982})}\BibitemShut {NoStop}%
\bibitem [{\citenamefont {Neuhauser}\ \emph {et~al.}(1991)\citenamefont
  {Neuhauser}, \citenamefont {Baer}, \citenamefont {Judson},\ and\
  \citenamefont {Kouri}}]{neuhauser1991application}%
  \BibitemOpen
  \bibfield  {author} {\bibinfo {author} {\bibfnamefont {D.}~\bibnamefont
  {Neuhauser}}, \bibinfo {author} {\bibfnamefont {M.}~\bibnamefont {Baer}},
  \bibinfo {author} {\bibfnamefont {R.~S.}\ \bibnamefont {Judson}}, \ and\
  \bibinfo {author} {\bibfnamefont {D.~J.}\ \bibnamefont {Kouri}},\ }\href@noop
  {} {\bibfield  {journal} {\bibinfo  {journal} {Comput. Phys. Commun.}\
  }\textbf {\bibinfo {volume} {63}},\ \bibinfo {pages} {460--481} (\bibinfo
  {year} {1991})}\BibitemShut {NoStop}%
\bibitem [{\citenamefont {Tully}(1990)}]{tully1990molecular}%
  \BibitemOpen
  \bibfield  {author} {\bibinfo {author} {\bibfnamefont {J.~C.}\ \bibnamefont
  {Tully}},\ }\href@noop {} {\bibfield  {journal} {\bibinfo  {journal} {J.
  Chem. Phys.}\ }\textbf {\bibinfo {volume} {93}},\ \bibinfo {pages}
  {1061--1071} (\bibinfo {year} {1990})}\BibitemShut {NoStop}%
\bibitem [{\citenamefont {Lawrence}\ and\ \citenamefont
  {Manolopoulos}(2019)}]{lawrence2019analysis}%
  \BibitemOpen
  \bibfield  {author} {\bibinfo {author} {\bibfnamefont {J.~E.}\ \bibnamefont
  {Lawrence}}\ and\ \bibinfo {author} {\bibfnamefont {D.~E.}\ \bibnamefont
  {Manolopoulos}},\ }\href@noop {} {\bibfield  {journal} {\bibinfo  {journal}
  {J. Chem. Phys.}\ }\textbf {\bibinfo {volume} {151}},\ \bibinfo {pages}
  {244109} (\bibinfo {year} {2019})}\BibitemShut {NoStop}%
\bibitem [{\citenamefont {Voth}, \citenamefont {Chandler},\ and\ \citenamefont
  {Miller}(1989)}]{voth1989rigorous}%
  \BibitemOpen
  \bibfield  {author} {\bibinfo {author} {\bibfnamefont {G.~A.}\ \bibnamefont
  {Voth}}, \bibinfo {author} {\bibfnamefont {D.}~\bibnamefont {Chandler}}, \
  and\ \bibinfo {author} {\bibfnamefont {W.~H.}\ \bibnamefont {Miller}},\
  }\href@noop {} {\bibfield  {journal} {\bibinfo  {journal} {J. Chem. Phys.}\
  }\textbf {\bibinfo {volume} {91}},\ \bibinfo {pages} {7749--7760} (\bibinfo
  {year} {1989})}\BibitemShut {NoStop}%
\bibitem [{\citenamefont {Predescu}\ and\ \citenamefont
  {Miller}(2005)}]{predescu2005optimal}%
  \BibitemOpen
  \bibfield  {author} {\bibinfo {author} {\bibfnamefont {C.}~\bibnamefont
  {Predescu}}\ and\ \bibinfo {author} {\bibfnamefont {W.~H.}\ \bibnamefont
  {Miller}},\ }\href@noop {} {\bibfield  {journal} {\bibinfo  {journal} {J.
  Phys. Chem. B}\ }\textbf {\bibinfo {volume} {109}},\ \bibinfo {pages}
  {6491--6499} (\bibinfo {year} {2005})}\BibitemShut {NoStop}%
\bibitem [{\citenamefont {Wigner}(1937)}]{wigner1937calculation}%
  \BibitemOpen
  \bibfield  {author} {\bibinfo {author} {\bibfnamefont {E.}~\bibnamefont
  {Wigner}},\ }\href@noop {} {\bibfield  {journal} {\bibinfo  {journal} {J.
  Chem. Phys.}\ }\textbf {\bibinfo {volume} {5}},\ \bibinfo {pages} {720--725}
  (\bibinfo {year} {1937})}\BibitemShut {NoStop}%
\bibitem [{\citenamefont {Wigner}(1938)}]{wigner1938transition}%
  \BibitemOpen
  \bibfield  {author} {\bibinfo {author} {\bibfnamefont {E.}~\bibnamefont
  {Wigner}},\ }\href@noop {} {\bibfield  {journal} {\bibinfo  {journal} {Trans.
  Faraday Soc.}\ }\textbf {\bibinfo {volume} {34}},\ \bibinfo {pages} {29--41}
  (\bibinfo {year} {1938})}\BibitemShut {NoStop}%
\bibitem [{\citenamefont {Herman}, \citenamefont {Bruskin},\ and\ \citenamefont
  {Berne}(1982)}]{herman1982path}%
  \BibitemOpen
  \bibfield  {author} {\bibinfo {author} {\bibfnamefont {M.}~\bibnamefont
  {Herman}}, \bibinfo {author} {\bibfnamefont {E.}~\bibnamefont {Bruskin}}, \
  and\ \bibinfo {author} {\bibfnamefont {B.}~\bibnamefont {Berne}},\
  }\href@noop {} {\bibfield  {journal} {\bibinfo  {journal} {J. Chem. Phys.}\
  }\textbf {\bibinfo {volume} {76}},\ \bibinfo {pages} {5150--5155} (\bibinfo
  {year} {1982})}\BibitemShut {NoStop}%
\end{thebibliography}%
\nocite{*}

\end{document}